\documentclass[
reprint,
aps,
pra,
superscriptaddress,
floatfix,
showkeys
]{revtex4-1}

\usepackage{
amsmath,
amssymb,
amsfonts,
graphicx,
setspace,
color,
soul,
hyperref,
upgreek,
bm,
dcolumn,
epstopdf,
picture,
float,
subfigure,
chngcntr,
cases,
soul
}
\counterwithin*{paragraph}{section}
\counterwithin*{paragraph}{subsection}
\soulregister\cite7
\soulregister\ref7
\soulregister\pageref7

\newcommand{\rvec}{\mathbf{r}}

\newcommand{\fvec}{\mathbf{f}}
\newcommand{\qvec}{\mathbf{q}}
\newcommand{\tauvec}{\mathbf{\tau}}

\newcommand{\Pvec}{\mathbf{P}}
\newcommand{\Evec}{\mathbf{E}}
\newcommand{\Rmat}{\mathbf{R}}
\newcommand{\pvec}{\mathbf{p}}

\newcommand{\uvec}{\mathbf{\hat u}}

\newcommand{\uz}{\mathbf{\hat z}}
\newcommand{\Ximat}{\mathbf{\Xi}}
\newcommand{\Mmat}{\mathbf{M}}
\newcommand{\Imat}{\mathbf{I}}
\definecolor{blue}{rgb}{.03,.27,.49}
\definecolor{green}{rgb}{0.5, 0.5, 0.0}
\definecolor{red}{rgb}{.86,.08,.24}

\begin{document}

\title{
Cooling the optical-spin driven limit cycle oscillations of a levitated gyroscope
}

\author{Yoshihiko Arita}
\affiliation{SUPA, School of Physics \& Astronomy, University of St Andrews, North Haugh, St Andrews KY16 9SS, United Kingdom}
\affiliation{Molecular Chirality Research Centre, Chiba University, 1-33 Yayoi-cho, Inage-ku, Chiba-shi 263-0022, Japan}

\author{Stephen H. Simpson}
\affiliation{Institute of Scientific Instruments of the Czech Academy of Science, v.v.i., Kr\'alovopolsk\'a 147, 612 64 Brno, Czech Republic}

\author{Graham D Bruce}
\affiliation{SUPA, School of Physics \& Astronomy, University of St Andrews, North Haugh, St Andrews KY16 9SS, United Kingdom}

\author{Ewan M. Wright}
\affiliation{College of Optical Sciences, The University of Arizona, Tucson, Arizona 85721-0094, USA}
\affiliation{SUPA, School of Physics \& Astronomy, University of St Andrews, North Haugh, St Andrews KY16 9SS, United Kingdom}

\author{Pavel Zem\'{a}nek}
\affiliation{Institute of Scientific Instruments of the Czech Academy of Science, v.v.i., Kr\'alovopolsk\'a 147, 612 64 Brno, Czech Republic}

\author{Kishan Dholakia}
\affiliation{SUPA, School of Physics \& Astronomy, University of St Andrews, North Haugh, St Andrews KY16 9SS, United Kingdom}
\affiliation{Molecular Chirality Research Centre, Chiba University, 1-33 Yayoi-cho, Inage-ku, Chiba-shi 263-0022, Japan}
\affiliation{College of Optical Sciences, The University of Arizona, Tucson, Arizona 85721-0094, USA}
\affiliation{Department of Physics, College of Science, Yonsei University, Seoul 03722, South Korea}
\affiliation{School of Biological Sciences, The University of Adelaide, Adelaide, South Australia, Australia}


\date{\today}

\begin{abstract}
\noindent The non-conservative, azimuthal forces associated with inhomogeneous optical-spin angular momentum play a critical role in optical trapping. Intriguingly, birefringent microspheres can be stably levitated and rapidly rotated in circularly polarized optical traps in ultra-high vacuum whereas isotropic spheres are typically destabilized and expelled, even at relatively modest pressures. Here we show that the resolution of this apparent key paradox rests in the form of the orientationally averaged, effective forces acting on the spinning birefringent particle. In particular, the effective azimuthal component is heavily suppressed and highly non-linear. As a consequence, non-conservative effects are strongly, if imperfectly, inhibited. Their influence is apparent only at very low pressures where we observe the formation of noisy, nano-scale limit cycles or orbits. Finally, we show how parametric feedback can synthesize a form of dissipation, necessary to preserve limit cycle oscillation, without introducing additional thermal fluctuations. This allows the preparation of highly coherent, self-sustained oscillations with effective temperatures on the order of a milliKelvin.
The tailoring of azimuthal spin forces through the material structure of a spinning, non-spherical particle opens up new opportunities for the design of ultra stable optical rotors. In addition, we have shown that the unique profile of the azimuthal force, featured in this work, allows for the formation of nano-scale limit cycles that can be stabilized and cooled. In principle, this approach could enable the cooling of limit cycles into the quantum regime, allowing for experimental realisation of quantum synchronization, or alternative ways of entangling mesoscopic bodies.

\end{abstract}


\maketitle

\addtocontents{toc}{\protect\setcounter{tocdepth}{-1}}


\section{Introduction}\label{sec:intro}
\noindent Levitated optomechanics based on optical forces relies on the use of mesoscopic particles suspended in vacuum using tightly focused laser light. The confinement and translational motion of such particles has seen extensive study in the last decade with major advances including parametric feedback cooling, zeptonewton force sensing and the realisation of cooling to the ground state of motion \cite{delic2020cooling,ranjit2016zeptonewton,gieseler2012subkelvin}.
\noindent Hand-in-hand with these advances has been the study of the rotational degree of freedom where the levitated particle’s transverse motion is not only confined, but the particle is also free to spin about its centre of mass. Circularly polarised light possesses spin angular momentum. A beam of such polarisation can stably trap and continuously rotate mesoscopic particles: by reducing the ambient pressure in a levitated geometry in vacuum, extraordinarily high spin rates may be available when operating micron-sized birefringent particles in vacuum~\cite{arita2013laser}. By extending to smaller particles, rotation rates of several GHz have recently been demonstrated~\cite{ahn2018optically,reimann2018ghz,ahn2020ultrasensitive,jin20216}. These achievements provide unprecedented access to a relatively unexplored physical regime. This may allow experiment to explore theoretically postulated quantum rotational effects, including quantum friction ~\cite{manjavacas2010vacuum,zhao2012rotational,stickler2018rotational,stickler2021quantum}. To progress this field further requires insight into the complex structure of the forces and torques acting on rapidly rotating particles in optical vacuum traps and the subsequent driven, stochastic motion.

\noindent A birefringent particle in a circularly polarized vacuum trap exhibits high stability in motion compared to a birefringent particle in a linearly polarized trap or an isotropic particle in a circularly polarized trap \cite{Svak2018Trans,arita2020coherent}. Our study explains the reasons for this enhanced stability. It is based on the role of azimuthal spin forces (ASFs).
For a birefringent sphere, we show that this force is orientation dependent and can even reverse its direction, so that it is directed against the incident momentum. For a rapidly spinning particle the effective (i.e. rotationally averaged) ASF is heavily suppressed, and locally non-linear, increasing the trap stability relative to that of an isotropic microsphere.\\
\noindent Although greatly reduced, the residual ASF is sufficient to push the centre of mass motion of the particle well beyond equilibrium. The associated effects, which include stochastic orbital rotation and the subsequent formation of noisy, nano-scale limit cycles, become increasingly conspicuous with decreasing pressure. Limit cycles, such as those observed here, are isolated periodic trajectories whose existence depends on a balance between the energy entering the system, in this case through the non-conservative forces, and the energy dissipated, in this case through viscous drag. A significant result of this paper concerns the application of parametric feedback cooling (FBC) to these noisy limit cycles. Here, FBC involves modulating the optical forces in time, in such a way as to synthesise effective dissipative forces. This FBC-induced damping is sufficient to preserve the limit cycle, without introducing additional thermal fluctuations. Decreasing the ambient pressure reduces intrinsic thermal fluctuations, leaving a coherent, cooled limit cycle with an effective temperature on the order of a few milliKelvin.\\ 

\noindent We note that our observations are generic to all optically levitated, light driven rotors. Our insights clarify the inherent stability of rotating birefringent particles in vacuum and pave the way for designing ultra-stable rotors capable of operating at higher optical power or lower pressure, especially those carrying the greater centripetal loads required to test fundamental material properties \cite{schuck2018ultrafast,gonzalez2021levitodynamics}. In particular, we have shown that the azimuthal forces, which necessarily destabilize circularly polarized vacuum traps, can be suppressed by the structure of the particle: careful engineering of optical rotors could eliminate this form of instability, allowing for ever greater spin rates.
Furthermore, we have demonstrated feedback cooling of limit cycle oscillations, developments of which could allow experimental realisation of quantum synchronization, macroscopic entanglement of mesoscopic bodies or, more generally, the study of the non-equilibrium stochastic thermodynamics of self-sustained oscillators \cite{walter2015quantum,roulet2018quantum,kato2019semiclassical,wachtler2019stochastic}.


\section{Results}\label{results}
\subsection{Overview of the experiment}

\begin{figure}[htb!]
\centering
\includegraphics[width = 1\columnwidth, clip = true]{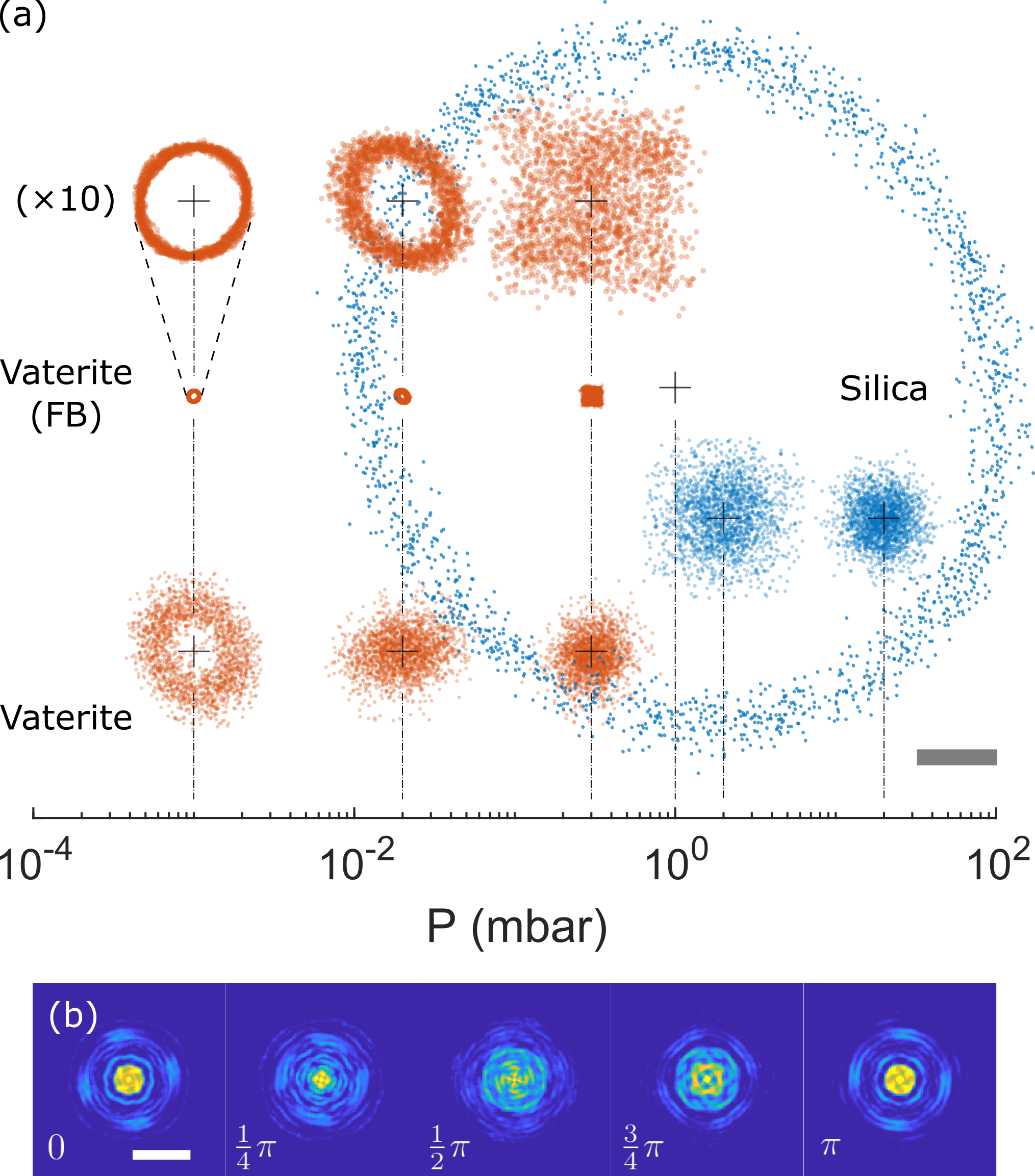}
\caption{
Overview of the experiment: 
(a) Scatter plots of the x-y coordinates of the centre of mass of  silica (blue) and vaterite (orange) microspheres before and after the Hopf bifurcation, without and with feedback (FB) cooling. The scale bar represents $200\,\mathrm{nm}$ and the centres of the distributions (marked as crosses) are positioned above the corresponding pressure, shown on the log scale below. (b) Intensity of light backscattered from a vaterite microsphere at a range of discrete orientations. This is used to track the centre of mass motion. 
}
\label{fig:overview}
\end{figure}
\noindent Circularly polarized light beams carry azimuthal components of momentum that swirl around their axes \cite{bekshaev2011internal}. The associated optical forces are necessarily non-conservative, pushing the trap out of equilibrium \cite{Svak2018Trans}. For isotropic microspheres, this results in striking and characteristic behaviour. With decreasing pressure, the stable trapping point undergoes a Hopf bifurcation \cite{simpson2021stochastic}, giving way to noisy limit cycles (or \textit{orbits})
whose amplitude increases until the particle is ultimately ejected from the trap. Intriguingly, similar behaviour has not been observed for birefringent, vaterite particles which are known to remain stably trapped even in ultra-high vacuum \cite{arita2013laser}.\\

\noindent Here we experimentally investigate this apparent discrepancy by analysing the centre of mass motion of a spinning vaterite particle in a circularly polarized trap under conditions of decreasing pressure. The optical trapping set-up is described in the Methods. The study makes use of two key techniques. First, the position of the centre of mass is recorded by using a quadrant photodiode (QPD) to track the orientationally averaged scattering pattern of the light passing through the birefringent particle (see Fig. (\ref{fig:overview})b), Methods and Supplementary Note \ref{sec:si_anisotropy}). Second, knowledge of the particle coordinates allows us to apply parametric feedback cooling to the centre of mass motion. An overview of the key results is given in Figure (\ref{fig:overview}a). In summary, we uncover behaviour analogous to that observed for isotropic spheres, signifying the active role of azimuthal spin forces. In comparison to isotropic spheres, however, the pressure required to form noisy limit cycles is about three orders of magnitude lower, and the dimensions of the limit cycles are about one order of magnitude lower. Application of feedback cooling results in the formation of ultra-coherent, nano-scale limit cycle oscillations, with effective temperatures on the scale of milliKelivin.  \\
Below, we provide more detailed analysis of the free running, and feedback cooled systems.

\subsection{Free running experiments}
\begin{figure*}[htb!]
\centering
\includegraphics[width = 1.0\textwidth, clip = true]{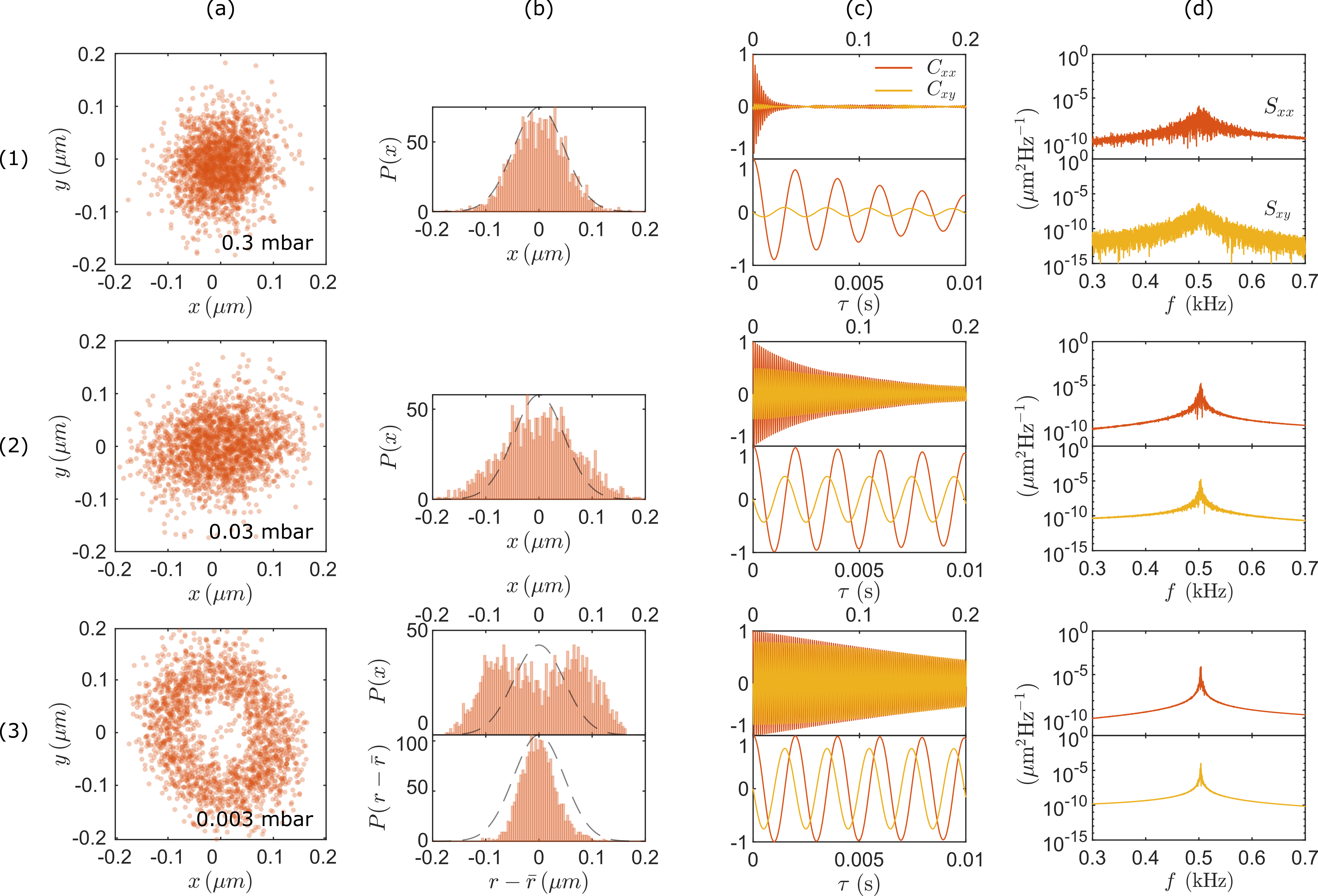}
\caption{
Experimentally measured dynamics of a vaterite microsphere trapped with a circularly polarised beam, showing the formation of limit cycle oscillations at low pressure. (a) Position distributions of the centre of mass in the $xy$ plane (transverse to the beam axis, $z$) and (b) their histograms in terms of $x$ in Cartesian coordinates and the radial $r$ position in radial coordinates, where $\bar{r}$ indicates the mean radial position. For comparison, the dashed lines show a Gaussian position distribution acquired at $10\,\mathrm{mbar}$. (c) Autocorrelation, $C_{xx}=\langle x(t)x(t+\tau) \rangle$ (orange) and cross correlation of $C_{xy}=\langle x(t) y(t+\tau) \rangle$ (yellow), where their decay times $\tau_{D}$ are $5.5\,\mathrm{ms}$, $16\,\mathrm{ms}$ ($0.3\,\mathrm{mbar}$); $41\,\mathrm{ms}$, $97\,\mathrm{ms}$ ($0.03\,\mathrm{mbar}$); $170\,\mathrm{ms}$, $180\,\mathrm{ms}$ ($0.003\,\mathrm{mbar}$), respectively. (d) Power spectral density of the $x$ coordinate, $S_{xx}$, and the cross power spectral density of $x$ and $y$, $S_{xy}$ showing the trap frequency at $f_x\approx f_y\sim0.5\,\mathrm{kHz}$. Rows (1) to (3) represent data at different gas pressures.
}
\label{fig:vaterite_cp_pmod_0_0_0__}
\end{figure*}

\noindent \paragraph{Description of results:} Figure~(\ref{fig:vaterite_cp_pmod_0_0_0__}) contains a detailed exploration of the centre of mass motion of a spinning vaterite microsphere. The radius of the microsphere is $a=2.2\mu$m, and the trap is a tightly focused circularly polarized beam of wavelength $1070$nm and numerical aperture $1.25$, see Methods for details, and the viscosity ($\mu$, in Pa s) varies linearly with pressure ($P$ in mbar) (Supplementary Note \ref{sec:si_detsims}). For a sphere of the given radius, 
\begin{equation}\label{eq:mup}
\mu \simeq 3.56 \times 10^{-7}P.
\end{equation}
For comparison, an analogous study, for a silica microsphere, is provided in Supplementary Note \ref{sec:si_isotropic}.
At higher pressures ($\approx 0.3$mbar), the centre of mass motion of spinning vaterite appears conservative (see top row, Fig. (\ref{fig:vaterite_cp_pmod_0_0_0__})). The spatial probability distribution function closely resembles the spatial part of the Boltzmann distribution for a particle in a parabolic potential i.e. it is normal, with kurtosis 2.99 (Fig. (\ref{fig:vaterite_cp_pmod_0_0_0__})1b). The $x$ and $y$ motions are only very weakly coupled, as indicated by the negligible amplitude of the cross correlation (Fig. (\ref{fig:vaterite_cp_pmod_0_0_0__})1c). As a consequence, the power spectral densities (PSDs) for motion in the $x$ and $y$ directions show resonant peaks at two slightly different frequencies, $\omega_x$ and $\omega_y$, separated by $\lesssim 10 \%$. As the pressure and, therefore, the viscous damping, is reduced the stochastic rotation of the centre of mass of the particle increases in amplitude and centripetal forces start to deform the probability distribution. Small departures from normality emerge at $\approx0.03$mbar, when the spatial distribution has a kurtosis of 2.9 (see, Fig. (\ref{fig:vaterite_cp_pmod_0_0_0__}),2b), developing into a pronounced, annular distribution with kurtosis 2.6 at around $0.003$mbar. That this transition is connected with driven orbital motion is clearly shown by the correlation functions (Fig. (\ref{fig:vaterite_cp_pmod_0_0_0__})2c and 3c) \cite{jones2009rotation}. In particular, the amplitude of the cross correlations rise sharply, from a negligible value at 0.3mbar, to $\approx 0.5$ at $0.03$mbar and $\approx 0.8$ at $0.003$mbar. Expanded figures show the relative phases of the auto-correlation $\sim \sin(\Omega_o t)e^{-t/\tau_D}$ and cross-correlation $\sim \cos(\Omega_o t)e^{-t/\tau_D}$ (for characteristic orbital frequency, $\Omega_o$ and decay times $\tau_D$) confirming that the orbits are approximately circular. The general increase in all exponential decay times with decreasing pressure represents an overall increase in coherence.  The growing tendency towards deterministic orbiting is further indicated by the coalescence of the decay times, $\tau_D$, for the correlation and cross correlation functions. The initial disparity, apparent at $0.3$mbar, indicates the presence of more than one weakly coupled process, while the single decay time apparent at $0.003$mbar suggests that a single, highly coherent process dominates. Finally, the power spectral densities (Fig. (\ref{fig:vaterite_cp_pmod_0_0_0__}) column d) show the transition from biased stochastic motion (i.e. a relatively weak tendency towards circulation), to a fluctuating, driven motion (i.e. fluctuations around an underlying, deterministic orbit or limit cycle). As the pressure is decreased the two spectral peaks, discernible at $0.3$mbar, merge, leaving a single, dominant frequency, corresponding to the fundamental frequency of an underlying limit cycle \cite{simpson2021stochastic}. This transition mirrors the behaviour of the decay constants of the correlation functions, providing further evidence that the two processes, resolvable at higher pressure, are replaced by a single, dominant, non-equilibrium process as damping is decreased.  \\
\noindent We note that the behaviour described above, for spinning vaterite micro-spheres, is qualitatively similar to that observed for silica micro-spheres (see \cite{svak2018transverse} and Supplementary Note \ref{sec:si_isotropic}). However, the quantitative differences are dramatic. For example, the critical pressure, necessary for limit cycle formation, is $\approx 0.003$mbar for vaterite and $\approx 1$mbar for silica. In addition, the orbit radius is $\approx 0.1\mu$m for the vaterite particle, compared with $\approx 1\mu$m for silica. These observations suggest that the azimuthal forces acting on the vaterite particle are small compared with those for silica and that the curvature in the force, necessary for limit cycle formation, is compressed into the region immediately surrounding the beam axis.\\

\begin{figure*}[htb!]
\centering
\includegraphics[width = 1\textwidth, clip = true]{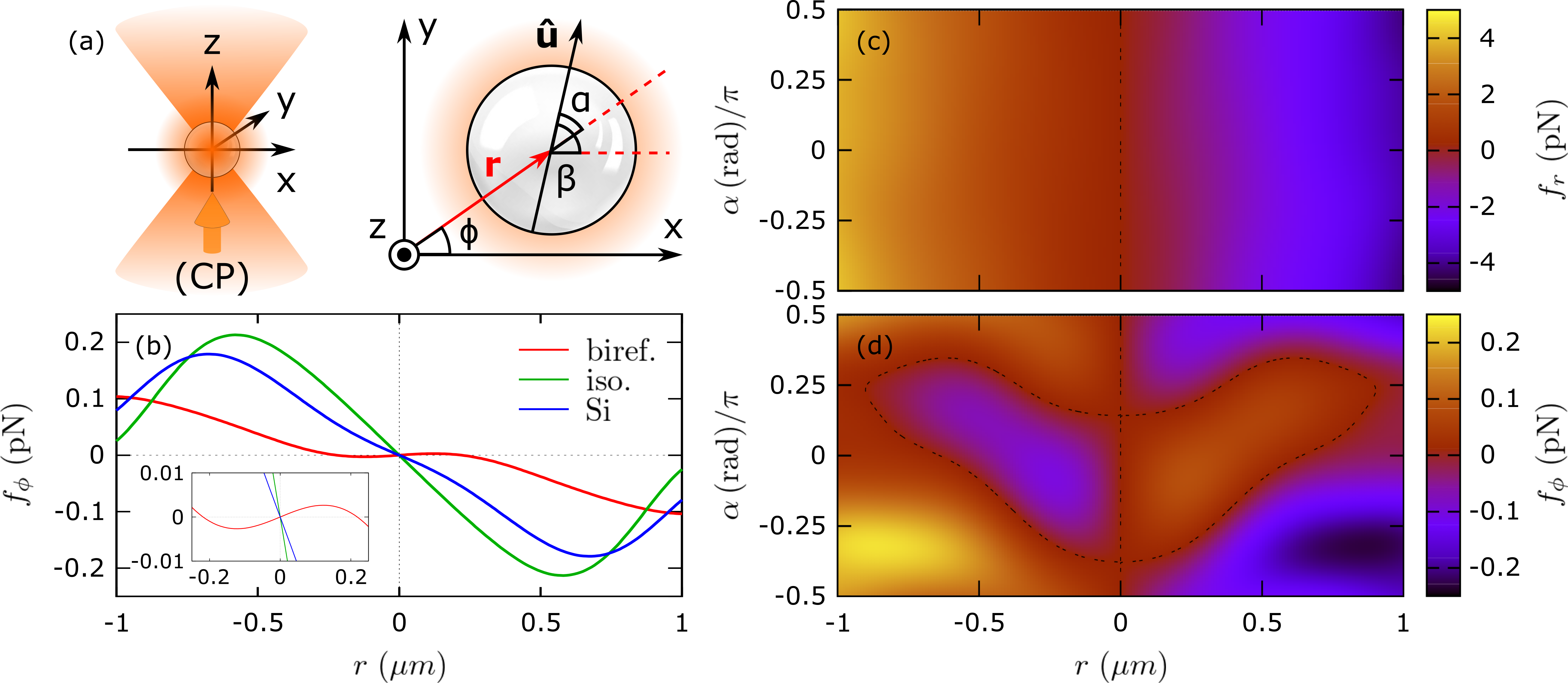}
\caption{
(a) Schematic showing the coordinate scheme used throughout. (b) Orientationally averaged ASFs acting on a birefringent microsphere (\textit{biref}), an equivalent isotropic microsphere (\textit{iso}) with the same mean refractive index, and a silica microsphere (\textit{Si}). The inset shows the same variations over an interval $-0.2 \mu m \leq x \leq 0.2 \mu m$. (c) The gradient force, $f_r(x,\alpha)$, and (d) transverse spin force, $f_\phi(x,\alpha)$ acting on a birefringent microsphere as a function of position and orientation. 
}
\label{fig:forces}
\end{figure*}

\paragraph{Theoretical model: } Insight into the huge quantitative differences between the motion of silica and vaterite micro-spheres can be obtained by considering a simple numerical model for the forces acting on a birefringent vaterite microsphere in an idealized, cylindrically symmetric, circularly polarized beam (see Supplementary Note \ref{sec:si_detsims}). The particle is at mechanical equilibrium when its symmetry axis, $\uvec$, is parallel to the transverse, $xy$ plane, and the centre of mass is downstream of the focal point, so that the weight of the particle is balanced by the upward radiation pressure (see Fig. (\ref{fig:forces})a). In this configuration, the particle experiences an optical torque, $\tau_z$, which causes it to spin.

\noindent Fig. (\ref{fig:forces})b-d shows the transverse forces acting on the sphere as a function of the radial displacement ($r$), and orientation ($\alpha$). The radial force, $f_r$, is very weakly dependent on orientation, and acts as a simple gradient force confining the particle in the trap, Fig. (\ref{fig:forces})c. In contrast, the azimuthal force, $f_\phi$, oscillates with varying $\alpha$ and can be well approximated by a low order Fourier expansion, $f_\phi(r,\alpha)\approx a_0(r) + a_2(r)\sin(2\alpha)+b_2(r)\cos(2\alpha)$, in which the $a_0$ is much smaller than $a_2$ or $b_2$, (Fig. (\ref{fig:forces})d). \\
The stochastic, dynamical motion produced by the forces described in Fig. (\ref{fig:forces})c,d, depends qualitatively on the relative time scales of the spinning rotation, and the translational motion of the vaterite. The equilibrium spin rate ($\Omega_s$) is given by balancing the optical spin torque ($\tau_z$) and the rotational drag, i.e. $\tau_z = \xi_r \Omega_s$, where $\xi_r=8\pi\mu a^3$ and $\mu$ is the viscosity and, as described further below, the translational frequency is independent of viscosity and pressure, $\Omega_o \approx \sqrt{k/m}$, where $k$ is the trap stiffness and $m$ the mass. At modest pressure, the two time scales are comparable, $\Omega_s \sim \Omega_o$. In this regime, spinning and translational motions of the particle interact, and the dynamics depends on the detailed force field of the system (see Supplementary Note \ref{sec:si_detsims}). Here we are primarily interested in the stability of the trap at very low pressure, where the particle spins very fast ($>100\,\mathrm{kHz}$). In this case $\Omega_s \gg \Omega_o$. Conservation of angular momentum increasingly confines the symmetry axis, $\uvec$, to the transverse ($xy$) plane and  fluctuations of the azimuthal force, caused by the particle spinning, are too rapid to couple with the motion of the centre of mass (Supplementary Note \ref{sec:si_detsims}). Under these conditions, the dynamical motion of the centre of mass is determined by rotationally averaged, effective forces, $\langle f_{r/\phi}(r) \rangle = \int^{2\pi}_0 f_{r/\phi}(r,\alpha)d\alpha$, the azimuthal component of which is strongly suppressed in comparison with isotropic particles (Fig. (\ref{fig:forces})b). The resulting Langevin equations of motion are,

\begin{subequations}\label{eq:Lang1}
\begin{align}
\langle \fvec(\rvec) \rangle +\fvec^L(t)-mg\uz-\xi_t \dot \rvec = m \ddot \rvec, \label{eq:Lang1a} \\
\langle \fvec^L(t) \rangle = 0, \;\; \langle \fvec^L(t) \otimes \fvec^L(t') \rangle = 2k_BT\xi_t\delta(t-t'), \label{eq:Lang1b}
\end{align}
\end{subequations}

\begin{figure}[htb!]
\centering
\includegraphics[width = 1\columnwidth, clip = true]{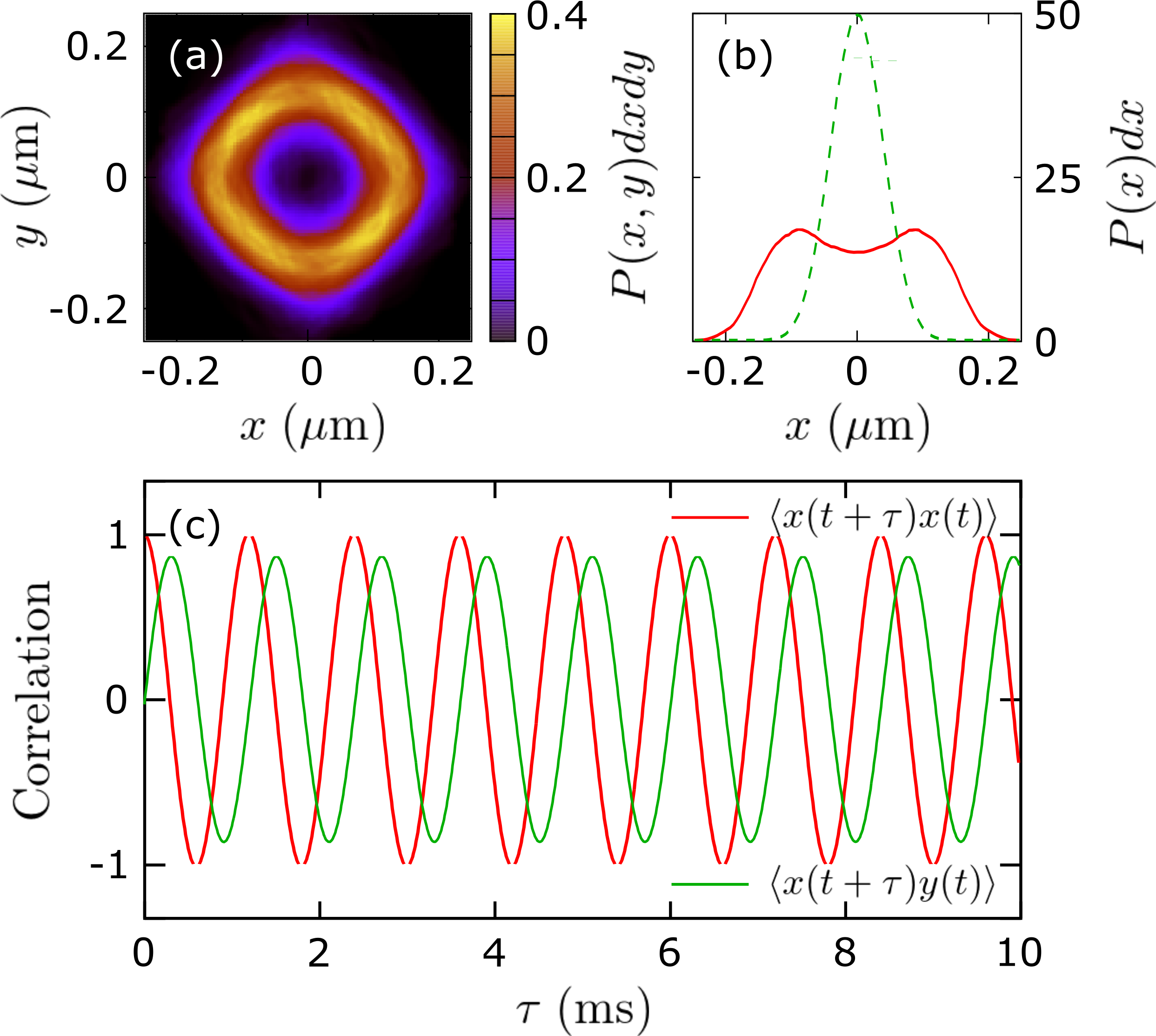}
\caption{
Spatial probability distribution function for a birefringent microsphere undergoing stochastic motion in an orientationally averaged force field in (a) two and (b) one spatial dimension. The autocorrelation, $\langle x(t+T)x(T)\rangle$, and cross correlation $\langle x(t+T)y(T)\rangle$ demonstrating coherent orbital motion, (c). }
\label{fig:sims}
\end{figure}

\noindent General features of this motion can be qualitatively understood as follows. For higher pressure (viscosity), the effective azimuthal force (ASF) biases the Brownian motion of the particle imparting a tendency towards orbital rotation about the beam axis. For higher pressures, the mean centripetal force is weaker than the dominant gradient force. As a result the trapping point remains stable and the spatial part of the probability density, caused by fluctuations about this fixed point, is approximately normal (with a kurtosis of 3). Reducing the drag (by reducing the pressure) allows the orbital angular momentum of the particle to grow. Eventually, the centripetal forces are sufficient to overcome the gradient forces and limit cycles form in which gradient forces balance centripetal forces, and azimuthal forces balance viscous drag, 

\begin{subequations}\label{eq:lceqm}
\begin{align}
mr\Omega_o^2 &=f_r(r) \approx kr,\label{eq:lceqma}\\
\xi_t \Omega_o r &= f_\phi(r) .\label{eq:lceqmb}
\end{align}
\end{subequations}

In addition to the equilibrium conditions, Eqs. (\ref{eq:lceqm}), stability requires that the forces have appropriate non-linearity (Supplementary Note \ref{sec:si_eqmorb}). For birefringent spheres, the required curvature appears in the effective ASF, Fig. (\ref{fig:forces})c, relatively close to the beam axis and within the linear range of the gradient force (justifying the final term on the right of Eq. (\ref{eq:lceqma})). Since $\Omega_o=\sqrt{k/m}$ is constant, equilibrium conditions for the limit cycle are determined by Eq. (\ref{eq:lceqmb}), and reduce to the intersection of the straight line, $\xi_t\Omega_o r$ with the curved, effective ASF, $\langle f_\phi(r) \rangle$ (Fig. (\ref{fig:forces})c). Importantly, as the drag is reduced, the radius of the orbit increases.\\
These equilibrium and stability conditions allow us to determine the range of viscosities for which the model will develop limit cycle oscillations. Fig. (\ref{fig:sims}) shows the results of stochastic simulations, numerically integrating Eq. (\ref{eq:Lang1}), see Supplementary Note \ref{sec:si_detsims}, with a viscosity of $\mu=2 \times 10^{-8}$Pas (equivalently, $P\approx$0.05mbar).
Consistent with the experiment, the model shows noisy limit cycles with a radius of $r_o\sim 0.1 \rightarrow 0.15 \mu$m, and a dramatically enhanced overall stability such that the particle remains in the trap for viscosities as low as $10^{-10}$Pas (see Supplementary Note \ref{sec:si_effsims}), or pressures of $\lesssim 3 \times 10^{-4}$mbar. Quantitative differences between the model and experiment have numerous causes. Most significantly, the calculation of the effective ASF is intrinsically inaccurate, since small errors in the absolute value of the forces can result in very large relative errors in the rotational average. For example, the average values of $\sin(2\alpha)+10^{-6}$ and $\sin(2\alpha)+10^{-7}$ differ by an order of magnitude, although the absolute values of the two functions are very close for most values of $\alpha$. Since the motion of the particle is extremely sensitive to the form of the effective ASF, the level of agreement with experiment may be considered remarkable from this perspective.

\subsection{Parametric feedback cooling}
\begin{figure*}[htb!]
\centering
\includegraphics[width = 1.0\textwidth, clip = true]{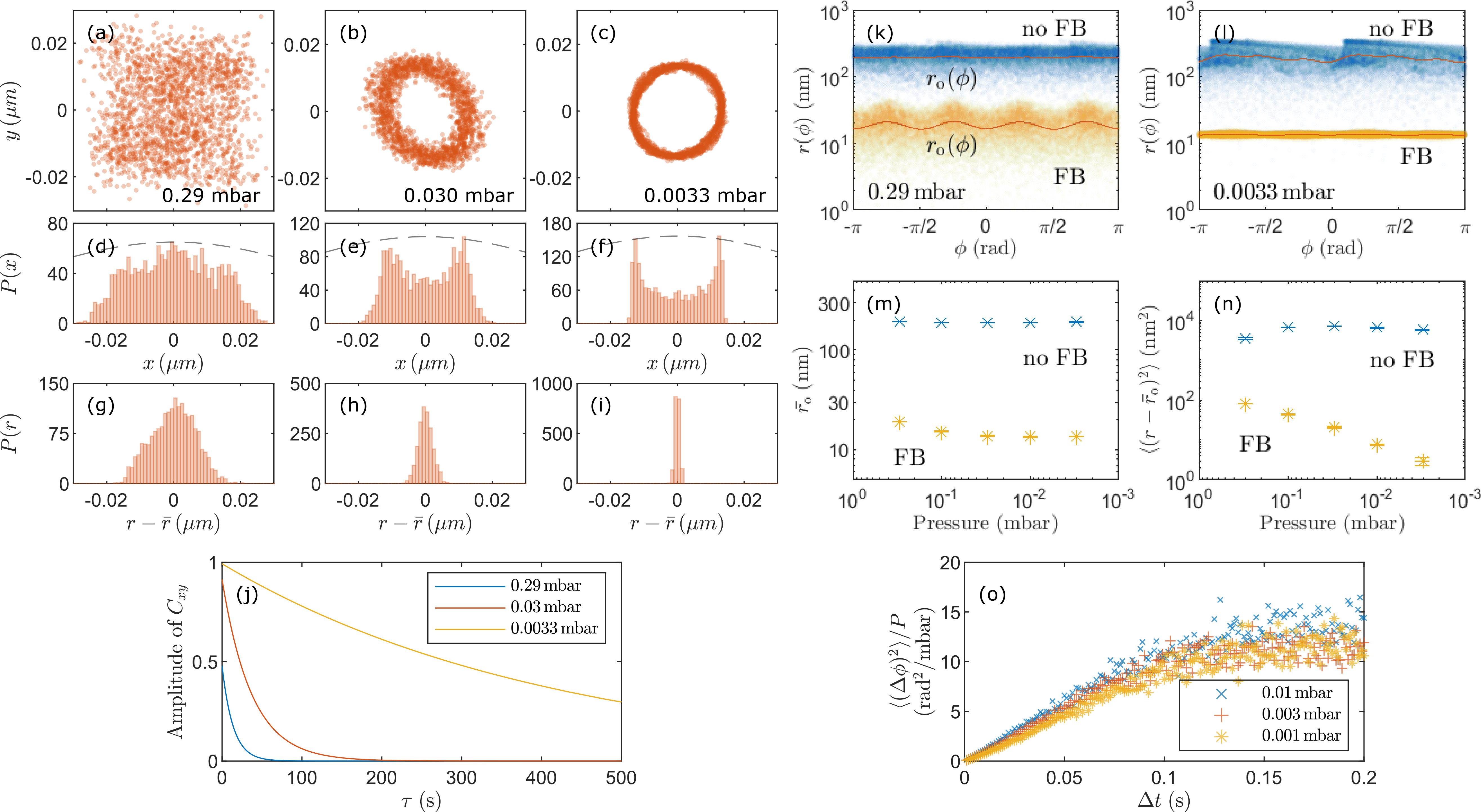}
\caption{
Stochastic trajectory of a vaterite microsphere in a circularly polarised trap with feedback damping. Particle position distribution in the $xy$ plane at different gas pressures of (a) $0.29\,\mathrm{mbar}$, (b) $0.03\,\mathrm{mbar}$, (c) $0.0033\,\mathrm{mbar}$ and their position histograms (d-f) in terms of $x$ in Cartesian coordinates and (g-i) $r$ in radial coordinates, where $\bar{r}$ indicates the mean radial position. The dashed lines in (d-f) show the Gaussian position distribution without feedback, at $10\,\mathrm{mbar}$. Envelope of the cross correlation $\langle x(t)y(t+\tau)\rangle$ at different gas pressures with decay times of $13.8$s ($0.28\,\mathrm{mbar}$), $37$s ($0.03\,\mathrm{mbar}$) and $416$s ($0.0033\,\mathrm{mbar}$), compared with a orbital time period of $2$ms (j). Fitted limit cycles with and without FB at 0.29mbar (k) and 0.0033mbar (l), mean orbital radius $\bar{r_o}$,  (m), and (n) variance of radial fluctuations$\langle (r-r_\mathrm{o})^2\rangle$.
Phase diffusion of feedback cooled limit cycles accrued over time intervals, $\Delta t$, normalised by gas pressure in mbar. Results for three different pressures are shown, $P=0.01,0.003,0.001$ mbar (o). 
}
\label{fig:parametric_feedback_}
\end{figure*}

\noindent \paragraph{Theoretical considerations: } Conceptually, the aim of parametric feedback cooling (FBC) is to augment the intrinsic viscous forces in a stochastic system (e.g. Eq. (\ref{eq:fb0a})), without modifying the variance of the fluctuating forces. Obviously, this modified system does not satisfy the fluctuation-dissipation theorem (Eq. (\ref{eq:Lang1b}). However, by applying the following transformations to Eq. (\ref{eq:Lang1}),
\begin{subequations}\label{eq:fb0}
\begin{align}
\xi_t &\rightarrow \xi_t'=(\xi_t + \xi_t^{fb}), \label{eq:fb0a}\\
T &\rightarrow T'=T\xi_t/(\xi_t + \xi^{fb}_t) \label{eq:fb0b},
\end{align}
\end{subequations}
(where $\xi_t^{fb}$ is a feedback induced drag coefficient) we see that the system with FB is equivalent to a new, effectively autonomous system which satisfies the fluctuation-dissipation theorem with increased drag, Eq. (\ref{eq:fb0a}), and a rescaled temperature, Eq. (\ref{eq:fb0b}).\\
\noindent Synthesis of additional viscous drag, as in Eq. (\ref{eq:fb0a}), can be achieved experimentally by weakly modulating the systematic forces in response to measurements of the system configuration. The attainable value of $\xi^{fb}_t$ depends on the efficiency and accuracy with which this can be achieved. Typically, $\xi^{fb}_t$ increases gradually as the magnitude of the thermal fluctuations decrease (i.e. as pressure decreases), reaching a limiting value imposed by the finite time scales in the experimental set-up. Our implementation of FBC is described in detail in the Methods. In summary, the centre of mass motion of the particle is tracked with a QPD (Supplementary Note \ref{sec:si_anisotropy}). At a moderate pressure an appropriate set of motional frequencies are selected. These frequencies are tracked with a phase locked loop (PLL) and the optical power in the trapping beam is modulated accordingly. The pressure in the vacuum chamber is steadily reduced and the dynamical motion of the particle is recorded. As well as introducing effective drag terms, force modulation modifies average forces. Providing the modulation is weak, this effect is relatively minor.\\
Feedback cooling is most commonly applied to systems with conservative forces. As a consequence, the steady state statistics are independent of viscous drag (since it is absent from the Boltzmann distribution), and can be understood purely in terms of the effective temperature, Eq. (\ref{eq:fb0b}). The simplest example is that of a linearly polarized Gaussian trap \cite{gieseler2012subkelvin}, where the forces are locally linear and conservative. The motion is described in terms of discrete, orthogonal modes and associated Eigen-frequencies which can be cooled independently. For each mode, the variance is, for example, $\langle x^2 \rangle = k_BT/k$, where $k$ is the stiffness, where $T$ is replaced by the effective temperature, $T'$ in Eq. (\ref{eq:fb0b}) when the feedback is applied, $\langle x'^2 \rangle = k_BT'/k$. Therefore, in this very special case, the effective temperature can be measured through the ratio of the variances, 
\begin{equation}\label{eq:Tefflcon}
\frac{T'}{T}=\frac{\langle x'^2 \rangle}{\langle x^2 \rangle}
\end{equation}
where $\langle x'^2 \rangle$ is the variance of the cooled trap.\\
\noindent For non-conservative systems, such as those studied here, the influence of FBC is more involved  \cite{gieseler2015non}. In general, the steady state distribution functions of non-equilibrium systems in the underdamped regime depend on viscous drag (e.g. \cite{Svak2018Trans,arita2020coherent}), sometimes critically, and both transformations in Eq. (\ref{eq:fb0}) are required. Except in particular cases, the forms of these distributions cannot be known \textit{a priori}, so the influence of FBC cannot be intuitively understood in terms of an effective temperature alone, as it can be for conservative systems.\\
\noindent \paragraph{Experimental results: } With these considerations in mind, we explore the effect of FBC on the non-equilibrium centre of mass motion of our rapidly spinning vaterite particles. Rather than having discrete orthogonal modes with distinct eigenvalues we have biased stochastic rotation (at higher pressure) or noisy limit cycles (at lower pressure). Figure (\ref{fig:parametric_feedback_}) describes the effect of applying FBC simultaneously to the $x$, $y$ and $z$ motions of optically trapped spinning vaterite microspheres. 
Fig. (\ref{fig:parametric_feedback_}),(a-i) shows various projections of the spatial probability distribution function (PDF). Decreasing pressure results in dramatic clarification of the underlying limit cycle, as illustrated by the cross correlations, Fig. (\ref{fig:parametric_feedback_}),j. This observation is further quantified in Fig. (\ref{fig:parametric_feedback_})(k-n). Deterministic limit cycles are approximated by fitting general closed curves, $r_o(\phi)$, to the noisy data (Supplementary Note \ref{sec:si_lcfit}) allowing us to compute first the mean radius of the cycle, $\bar{r_o}=\frac{1}{2\pi}\int^\pi_{-\pi}r_o(\phi)d\phi$, (Fig. (\ref{fig:parametric_feedback_})m) and next the variance of the fluctuations away from the cycle (Fig. (\ref{fig:parametric_feedback_})n). Blue and orange points in Fig. (\ref{fig:parametric_feedback_})m show that the dimensions of these FB cooled limit cycles decrease slowly from 1 mbar, tending towards a limiting value at lower pressures. The power spectral densities of these oscillations, not shown here, feature a single sharp peak at the fundamental frequency, $\Omega_o$, of the limit cycle which takes a value of $\approx 500$Hz, and does not vary significantly with pressure.\\
The parameters of these limit cycles can be used to quantify the effective forces operating in the FB cooled system. Balancing radial forces, Eq. (\ref{eq:lceqma}), gives the limit cycle frequency, $\Omega_o \approx \sqrt{k_{fb}/m}$, where $k_{fb}$ is the stiffness including any modifications caused by feedback. Since $\Omega_o$ is unchanged, the influence of FB on the trap stiffness is negligible (i.e. $k_{fb}\approx k$). Eq. (\ref{eq:lceqmb}) describes the relationship between the limit cycle dimensions and the drag forces acting on the particle. The gradual changes in limit cycle radius with pressure, Fig. (\ref{fig:parametric_feedback_})m, suggest that the total drag, $\sim-(\xi_t+\xi^{fb}_t)v_\phi$ increases to a limiting value as the pressure is reduced and $\xi_t$ becomes negligible. Thereafter the process is dominated by the FB term, $-\xi^{fb}_tv_\phi$, which continues to increase slightly with decreasing pressure, as the FB cooling becomes more efficient, before saturating.

\noindent While the dimensions of the FB cooled limit cycles slowly approach a limit, the variances of the cooled radial fluctuations decrease in proportion to the pressure, P i.e. $\langle(r-\bar{r})^2\rangle \propto P$, Fig. (\ref{fig:parametric_feedback_})n. Eqs. (\ref{eq:fb0}) cannot be applied directly, since we do not know the complete distribution function. However, observed linearity in the pressure dependence implies that Eq. (\ref{eq:Tefflcon}) is a meaningful measure of effective temperature, suggesting cooling by $\sim 4$ orders of magnitude at a pressure of $0.0033$mbar.


\noindent Further insight into the stochastic dynamics can be obtained by considering the phase diffusion about the limit cycle. Since limit cycles are neutrally stable, diffusion along a limit cycle resembles diffusion in a constant force field \cite{van1992stochastic,pikovsky2002synchronization}. For an approximately circular limit cycle with radius $\bar{r_o}$, at long times, the variance of the displacement of the phase satisfies Eq. (\ref{eq:pdiffa}),

\begin{subnumcases}
{\label{eq:pdiff}\langle \big( \phi(\Delta t) - \bar{\phi}(\Delta t) \big)^2 \rangle \sim}
\frac{2k_BT}{\bar{r_o}^2\xi_t}\Delta t {\label{eq:pdiffa}}\\
\frac{2k_BT\xi_t}{\bar{r_o}^2(\xi^{fb}_t+\xi_t)^2}\Delta t {\label{eq:pdiffb}}
\end{subnumcases}
\noindent Hypothesising, again, that FBC modifies the translational friction according to Eq. (\ref{eq:fb0a}) without changing the variance of the stochastic forces we anticipate that FB cooled phase diffusion follows Eq. (\ref{eq:pdiffb}), which reduces to the standard result, for a system without FB, when $\xi_t^{fb}=0$. We note that Eq. (\ref{eq:pdiffb}) can be obtained either by formal integration (Supplementary Note \ref{sec:si_phasediff}), or by applying the transformations, Eq. (\ref{eq:fb0}), directly to Eq, (\ref{eq:pdiffa}). In the low pressure limit we have $\langle \big( \phi(\Delta t) - \bar{\phi}(\Delta t) \big)^2 \rangle \propto \xi_t \Delta t$. This relationship is confirmed in Fig. (\ref{fig:parametric_feedback_})o, for time intervals $t \lesssim 0.1$s, and should be compared with the result for unperturbed limit cycles, Eq. (\ref{eq:pdiffa}), which has the inverse dependence on $\xi_t$. For larger intervals, $\Delta t \gtrsim 0.1$, phase diffusion saturates, suggesting that the particle motion, along the limit cycle, synchronizes with the time modulation of the optical forces \cite{pikovsky2000phase}. These observations support the notion that the statistics of FB cooled, non-equilibrium states can be understood in terms of the effective temperature and modified drag given in Eq. (\ref{eq:fb0}), given prior knowledge of the general form of the statistics we are interested in (Eq. (\ref{eq:pdiff}), for example).



\section{Discussion}
\noindent We have analysed the detailed stochastic motion of a birefringent, vaterite microsphere in a circularly polarized optical vacuum trap with and without parametric feedback cooling. The work has two significant and inter-related conclusions which we discuss separately below.\\ 
\paragraph{Enhanced stability: } First, we have resolved an apparent paradox relating to the relative stability of birefringent and isotropic microspheres in circularly polarized optical vacuum traps. As is now well understood, circularly polarized beams carry azimuthal components of momentum that swirl about the beam axis \cite{bekshaev2011internal,Svak2018Trans}. For isotropic microspheres in this regime, the associated non-conservative forces are sufficient to destabilize the trap when the ambient pressure falls below a moderate threshold (here, $\approx 1$mbar). Our results show that the greatly enhanced stability of birefringent spheres is caused by the surprising way in which they couple to azimuthal momentum. In particular, the amplitude of the azimuthal force acting on a birefringent particle oscillates as its orientation varies. When rapidly spinning, the centre of mass moves in response to rotationally averaged azimuthal forces which, for small displacements from the beam axis, are greatly suppressed in comparison with those acting on an isotropic sphere with the same mean refractive index. We emphasise that, for particular orientations, the azimuthal force acts in the opposite direction to the azimuthal momentum. Although counter-intuitive, such apparent violations of momentum conservation are abundant in opto-mechanics. Linear and angular tractor beams, in which optical forces oppose optical momentum flows, have been extensively studied \cite{magallanes2018macroscopic,brzobohaty2013experimental,han2018crossover,simpson2007optical}.  Of more immediate relevance are the observations of azimuthal force reversal made by Diniz et al, \cite{diniz2019negative}, who observed the reversal in direction of the azimuthal force acting on isotropic spheres of varying size and index. In each case, the anomalous behaviour can be accounted for by considering the momentum carried by the scattered field, in addition to that of the incident field. For forces connected with inhomogeneous spin, and associated momentum components, restrictions on the transverse momentum of the scattered light are particularly loose, explaining the diverse range of mechanical effects \cite{bliokh2014extraordinary,antognozzi2016direct}.

\noindent These principles are generic. Since they depend on azimuthal components of optical momentum that are intrinsic features of circularly polarized beams they will influence the stability of any object held in a circularly polarized trap, remaining significant in the small particle limit (Supplementary Note \ref{sec:si_scaling}). The observed suppression of azimuthal force, for vaterite particles, suggests that particles or beams could be engineered to achieve still greater suppression and therefore stability. This strategy could enable the stable rotation of large objects carrying the high centripetal loads required for testing fundamental material properties \cite{schuck2018ultrafast,gonzalez2021levitodynamics}. 

\noindent \paragraph{Parametric feedback cooling of limit cycle oscillations: } The second conclusion of this work concerns parametric feedback cooling of non-equilibrium steady states. This technique is most commonly applied to conservative, linear systems where the stochastic dynamics are relatively simple \cite{gieseler2012subkelvin}. Here, we apply it to noisy limit cycles induced by non-conservative azimuthal spin forces. The resulting dynamics can be described in terms of an equivalent autonomous system with drag forces increased by an additive feedback term, $\xi'\sim(\xi_t + \xi^{fb}_t)$, and operating at a reduced effective temperature, $T'\sim T\xi_t/(\xi_t + \xi^{fb}_t)$. This principle is consistent with the observed scaling behaviour of the limit cycle dimensions, the radial fluctuations and the phase diffusion along the cycle. In particular, the dimensions of the feedback cooled limit cycles are controlled by the drag experienced by the particle. While the intrinsic viscous drag, due to motion through the ambient gas, decreases with decreasing pressure, the feedback induced contribution increases slightly as the feedback becomes more efficient. Eventually the effective drag is dominated by the feedback term which itself approaches a limit due to finite time constants in the experimental equipment. The dimensions of the limit cycle reflect this, decreasing slightly in radius, and approaching a limit for low pressures, $P\lesssim 0.003$ mbar.\\
Phase diffusion about the limit cycle is well understood in terms of the effective drag and temperature. For short times, $\Delta t \lesssim 0.1$s, the variance in the phase difference is $\propto \xi_t \Delta t$ (i.e. decreasing with decreasing pressure) with feedback, compared with 
$\propto \Delta t/\xi_t$ without feedback\cite{van1992stochastic}. For greater times, the diffusion saturates, suggesting that the particle motion synchronizes with the force modulation \cite{pikovsky2000phase,amitai2017synchronization}. Fluctuations transverse to the limit cycle are harder to understand rigorously, since we do not have explicit, closed form expressions for the required probability distribution function. However, linear approximations \cite{Svak2018Trans} suggest a variation $\propto T$, at low pressure when the drag has reached its limiting value, $\xi^{fb}_t$. This is supported by the observed pressure dependence of the radial fluctuations, which are consistent with cooling to effective milliKelvin temperatures. We note that this statement requires cautious interpretation: the effective temperature can be thought of as the temperature that the physical system would have to be cooled to, in order to suppress the fluctuations to a similar degree. Definitions of temperature for single particles, out of equilibrium, in non-conservative environments are subtle and controversial \cite{dieterich2015single,casas2003temperature}. The detailed dynamics of a feedback cooled, non-equilibrium system may not be identical to those of a physically cooled system, even when low order moments (e.g. variances) of the fluctuations are the same. This is exemplified by the phase diffusion results, Fig. (\ref{fig:parametric_feedback_})o which saturate for time intervals $\gtrsim 0.1$s: this behaviour cannot be explained in terms of an effective temperature, but indicates interaction between the detailed motion of the particle and the time variation of the force modulation. Nevertheless, our results show that dynamical attractors, other than stable fixed points, can be cooled in some sense. More generally, cooling of limit cycle oscillations, using parametric feedback or the techniques of cavity optomechanics \cite{aspelmeyer2014cavity}, could provide a route to cooling limit cycle oscillators into the quantum regime paving the way for experimental realisation of quantum synchronization of macroscopic particles \cite{walter2015quantum,kato2019semiclassical} and providing alternative mechanisms for the entanglement of macroscopic bodies \cite{witthaut2017classical,roulet2018quantum}. 

\section*{Methods}\label{sec:methods}
\parindent=0em\subparagraph{Sample preparation.} Vaterite is a positive uniaxial birefringent material in a spherical morphology. The synthesis of vaterite microspheres with a mean radius of $2.20\,\upmu\mathrm{m}\pm0.02\,\upmu\mathrm{m}$ ($2\sigma$) is reported elsewhere~\cite{arita2013laser}. NIST-traceable size standards of silica with a diameter of $5.1\pm0.5\,\upmu\mathrm{m}$ (Thermo Scientific 9005) are used to compare their dynamics with birefringent microspheres. 

\subparagraph{Sample loading.} We use a small vacuum chamber with a volume of $27.7\,\upmu\ell$ and an annular piezoelectric transducer (APC International Ltd., Cat. no.70-2221) attached to the chamber to load microspheres into the optical trap. Before conducting the trapping experiments, dried microspheres are applied to the surface of optical glass windows (Harvard Apparatus Ltd., CS-8R: $8\,\mathrm{mm}$ in diameter, $150\,\upmu\mathrm{m}$ in thickness) of the chamber. Once sealed, the chamber pressure is reduced to $\sim100\,\mathrm{mbar}$. The piezoelectric transducer is operated at $140\,\mathrm{kHz}$ to detach microspheres from the glass surface, while a high numerical aperture microscope objective (Nikon Ltd., E Plan 100$\times$, NA=1.25/oil) focuses a circularly polarised trapping beam (continuous wave $1070\,\mathrm{nm}$) in the vacuum chamber. When a single particle is trapped, the piezoelectric transducer is switched off, and the chamber pressure is further reduced to $<1\,\mathrm{mbar}$ to provide parametric feedback control. The optical power can be adjusted to $10-25\,\mathrm{mW}$ to obtain the desired trap frequency from $0.4\,\mathrm{kHz}$ to $1.1\,\mathrm{kHz}$. 

\parindent=0em\subparagraph{Particle position detection} To calibrate the QPD response to nanometer displacements, we used a nano-positioning stage (PI, P-733.3 XYZ) with a carefully orientated vaterite microsphere adhered to the surface of the glass coverslip. First, the trapping beam is focused onto the centre of the stuck microsphere. Next, the forward scattered light from the microsphere is directed onto the QPD, and its voltage reading is recorded at ten-nanometer increments along the $x$ and $y$ directions. Then, the measurement is repeated at different orientations from 0 to $\pi$ with a $\pi/8$ step. As a result, we obtain eight different values of the QPD voltage dependence with respect to nanometre displacement (see Supplementary Note \ref{sec:si_anisotropy},  Fig.~(\ref{fig:vaterite_cp_back_scattering_}). Because the vaterite microsphere rotates at a rate ($\gg10\,\mathrm{kHz}$) that is orders of magnitude larger than the trap frequencies ($\sim0.5\,\mathrm{kHz}$) for gas pressures $<0.1\,\mathrm{mbar}$, the QPD voltage response can be averaged over the angles. As a result, we obtain a mean dependence of $5.59\,\mathrm{mV\,nm^{-1}}$ with a position sensitivity of $2.0\,\mathrm{nm}$ in its linear range (see Fig.~\ref{fig:vaterite_cp_back_scattering_}(b)).

\subparagraph{Feedback control and phase-locked loop.} In order to perform feedback cooling experiments, the particle motion is tracked by a quadrant photodiode (QPD) array (First Sensor, QP50-6SD2, -3dB at $150\,\mathrm{kHz}$). The interference pattern of the forward scattered light from a trapped microsphere is projected at the back focal plane of an imaging objective onto the QPD, yielding three voltage signals corresponding to the microsphere's $x$, $y$ and $z$ motion (see calibration of QPD above). 

\parindent=1em The QPD signals are processed by a lock-in-amplifier (Zurich Instruments, HF2LI, $210\,\mathrm{MSa/s}$, $\mathrm{DC}-50\,\mathrm{MHz}$) to extract the oscillation frequencies ($\omega_x, \omega_y, \omega_z$) and their phases ($\phi_x,\phi_y,\phi_z$) of the particle oscillation. A frequency-doubled waveform with an adjusted phase shift relative to the particle oscillation for each direction is superimposed as $\Sigma A_i(2\omega_it+\phi_i+\delta\phi_i)$, where $i=x,y,z$ to the voltage waveform driving an acousto-optic modulator (IntraAction, DTD-274HD6M) to modulate the trap intensity ($\leq\pm 5\%$). For limit cycle oscillations, $\omega_x=\omega_y\equiv \Omega_o$ and the modulation in response to transverse motions is simply $A\sin(2\Omega_\mathrm{o}t+\phi_\mathrm{o} + \delta\phi_\mathrm{o})$.
\parindent=0em

\subparagraph{Acknowledgements} YA, GDB and KD acknowledge support from the UK Engineering and Physical Sciences Research Council (EP/P030017/1). KD acknowledges support from the Australian Research Council. SHS and PZ acknowledge financial support from the Czech Science Agency (19-17765S) and the Czech Academy of Sciences (Praemium Academiae). SHS further acknowledges financial support from 
Ministerstvo \v{S}kolstv\'{i}, Ml\'{a}de\v{z}e a T\v{e}lov\'{y}chovy (CZ.02.1.01/0.0/0.0/15\_003/0000476)


\subparagraph{Competing financial interests} The authors declare no competing financial interests. 

\parindent=1em

\bibliography{main}

\begin{thebibliography}{47}%
\makeatletter
\providecommand \@ifxundefined [1]{%
 \@ifx{#1\undefined}
}%
\providecommand \@ifnum [1]{%
 \ifnum #1\expandafter \@firstoftwo
 \else \expandafter \@secondoftwo
 \fi
}%
\providecommand \@ifx [1]{%
 \ifx #1\expandafter \@firstoftwo
 \else \expandafter \@secondoftwo
 \fi
}%
\providecommand \natexlab [1]{#1}%
\providecommand \enquote  [1]{``#1''}%
\providecommand \bibnamefont  [1]{#1}%
\providecommand \bibfnamefont [1]{#1}%
\providecommand \citenamefont [1]{#1}%
\providecommand \href@noop [0]{\@secondoftwo}%
\providecommand \href [0]{\begingroup \@sanitize@url \@href}%
\providecommand \@href[1]{\@@startlink{#1}\@@href}%
\providecommand \@@href[1]{\endgroup#1\@@endlink}%
\providecommand \@sanitize@url [0]{\catcode `\\12\catcode `\$12\catcode
  `\&12\catcode `\#12\catcode `\^12\catcode `\_12\catcode `\%12\relax}%
\providecommand \@@startlink[1]{}%
\providecommand \@@endlink[0]{}%
\providecommand \url  [0]{\begingroup\@sanitize@url \@url }%
\providecommand \@url [1]{\endgroup\@href {#1}{\urlprefix }}%
\providecommand \urlprefix  [0]{URL }%
\providecommand \Eprint [0]{\href }%
\providecommand \doibase [0]{http://dx.doi.org/}%
\providecommand \selectlanguage [0]{\@gobble}%
\providecommand \bibinfo  [0]{\@secondoftwo}%
\providecommand \bibfield  [0]{\@secondoftwo}%
\providecommand \translation [1]{[#1]}%
\providecommand \BibitemOpen [0]{}%
\providecommand \bibitemStop [0]{}%
\providecommand \bibitemNoStop [0]{.\EOS\space}%
\providecommand \EOS [0]{\spacefactor3000\relax}%
\providecommand \BibitemShut  [1]{\csname bibitem#1\endcsname}%
\let\auto@bib@innerbib\@empty
\bibitem [{\citenamefont {Deli{\'c}}\ \emph {et~al.}(2020)\citenamefont
  {Deli{\'c}}, \citenamefont {Reisenbauer}, \citenamefont {Dare}, \citenamefont
  {Grass}, \citenamefont {Vuleti{\'c}}, \citenamefont {Kiesel},\ and\
  \citenamefont {Aspelmeyer}}]{delic2020cooling}%
  \BibitemOpen
  \bibfield  {author} {\bibinfo {author} {\bibfnamefont {U.}~\bibnamefont
  {Deli{\'c}}}, \bibinfo {author} {\bibfnamefont {M.}~\bibnamefont
  {Reisenbauer}}, \bibinfo {author} {\bibfnamefont {K.}~\bibnamefont {Dare}},
  \bibinfo {author} {\bibfnamefont {D.}~\bibnamefont {Grass}}, \bibinfo
  {author} {\bibfnamefont {V.}~\bibnamefont {Vuleti{\'c}}}, \bibinfo {author}
  {\bibfnamefont {N.}~\bibnamefont {Kiesel}}, \ and\ \bibinfo {author}
  {\bibfnamefont {M.}~\bibnamefont {Aspelmeyer}},\ }\href@noop {} {\bibfield
  {journal} {\bibinfo  {journal} {Science}\ }\textbf {\bibinfo {volume}
  {367}},\ \bibinfo {pages} {892} (\bibinfo {year} {2020})}\BibitemShut
  {NoStop}%
\bibitem [{\citenamefont {Ranjit}\ \emph {et~al.}(2016)\citenamefont {Ranjit},
  \citenamefont {Cunningham}, \citenamefont {Casey},\ and\ \citenamefont
  {Geraci}}]{ranjit2016zeptonewton}%
  \BibitemOpen
  \bibfield  {author} {\bibinfo {author} {\bibfnamefont {G.}~\bibnamefont
  {Ranjit}}, \bibinfo {author} {\bibfnamefont {M.}~\bibnamefont {Cunningham}},
  \bibinfo {author} {\bibfnamefont {K.}~\bibnamefont {Casey}}, \ and\ \bibinfo
  {author} {\bibfnamefont {A.~A.}\ \bibnamefont {Geraci}},\ }\href@noop {}
  {\bibfield  {journal} {\bibinfo  {journal} {Phys. Rev. A}\ }\textbf {\bibinfo
  {volume} {93}},\ \bibinfo {pages} {053801} (\bibinfo {year}
  {2016})}\BibitemShut {NoStop}%
\bibitem [{\citenamefont {Gieseler}\ \emph {et~al.}(2012)\citenamefont
  {Gieseler}, \citenamefont {Deutsch}, \citenamefont {Quidant},\ and\
  \citenamefont {Novotny}}]{gieseler2012subkelvin}%
  \BibitemOpen
  \bibfield  {author} {\bibinfo {author} {\bibfnamefont {J.}~\bibnamefont
  {Gieseler}}, \bibinfo {author} {\bibfnamefont {B.}~\bibnamefont {Deutsch}},
  \bibinfo {author} {\bibfnamefont {R.}~\bibnamefont {Quidant}}, \ and\
  \bibinfo {author} {\bibfnamefont {L.}~\bibnamefont {Novotny}},\ }\href@noop
  {} {\bibfield  {journal} {\bibinfo  {journal} {Phys. Rev. Lett.}\ }\textbf
  {\bibinfo {volume} {109}},\ \bibinfo {pages} {103603} (\bibinfo {year}
  {2012})}\BibitemShut {NoStop}%
\bibitem [{\citenamefont {Arita}\ \emph {et~al.}(2013)\citenamefont {Arita},
  \citenamefont {Mazilu},\ and\ \citenamefont {Dholakia}}]{arita2013laser}%
  \BibitemOpen
  \bibfield  {author} {\bibinfo {author} {\bibfnamefont {Y.}~\bibnamefont
  {Arita}}, \bibinfo {author} {\bibfnamefont {M.}~\bibnamefont {Mazilu}}, \
  and\ \bibinfo {author} {\bibfnamefont {K.}~\bibnamefont {Dholakia}},\
  }\href@noop {} {\bibfield  {journal} {\bibinfo  {journal} {Nat. Commun.}\
  }\textbf {\bibinfo {volume} {4}},\ \bibinfo {pages} {2374} (\bibinfo {year}
  {2013})}\BibitemShut {NoStop}%
\bibitem [{\citenamefont {Ahn}\ \emph {et~al.}(2018)\citenamefont {Ahn},
  \citenamefont {Xu}, \citenamefont {Bang}, \citenamefont {Deng}, \citenamefont
  {Hoang}, \citenamefont {Han}, \citenamefont {Ma},\ and\ \citenamefont
  {Li}}]{ahn2018optically}%
  \BibitemOpen
  \bibfield  {author} {\bibinfo {author} {\bibfnamefont {J.}~\bibnamefont
  {Ahn}}, \bibinfo {author} {\bibfnamefont {Z.}~\bibnamefont {Xu}}, \bibinfo
  {author} {\bibfnamefont {J.}~\bibnamefont {Bang}}, \bibinfo {author}
  {\bibfnamefont {Y.-H.}\ \bibnamefont {Deng}}, \bibinfo {author}
  {\bibfnamefont {T.~M.}\ \bibnamefont {Hoang}}, \bibinfo {author}
  {\bibfnamefont {Q.}~\bibnamefont {Han}}, \bibinfo {author} {\bibfnamefont
  {R.-M.}\ \bibnamefont {Ma}}, \ and\ \bibinfo {author} {\bibfnamefont
  {T.}~\bibnamefont {Li}},\ }\href@noop {} {\bibfield  {journal} {\bibinfo
  {journal} {Phys. Rev. Lett.}\ }\textbf {\bibinfo {volume} {121}},\ \bibinfo
  {pages} {033603} (\bibinfo {year} {2018})}\BibitemShut {NoStop}%
\bibitem [{\citenamefont {Reimann}\ \emph {et~al.}(2018)\citenamefont
  {Reimann}, \citenamefont {Doderer}, \citenamefont {Hebestreit}, \citenamefont
  {Diehl}, \citenamefont {Frimmer}, \citenamefont {Windey}, \citenamefont
  {Tebbenjohanns},\ and\ \citenamefont {Novotny}}]{reimann2018ghz}%
  \BibitemOpen
  \bibfield  {author} {\bibinfo {author} {\bibfnamefont {R.}~\bibnamefont
  {Reimann}}, \bibinfo {author} {\bibfnamefont {M.}~\bibnamefont {Doderer}},
  \bibinfo {author} {\bibfnamefont {E.}~\bibnamefont {Hebestreit}}, \bibinfo
  {author} {\bibfnamefont {R.}~\bibnamefont {Diehl}}, \bibinfo {author}
  {\bibfnamefont {M.}~\bibnamefont {Frimmer}}, \bibinfo {author} {\bibfnamefont
  {D.}~\bibnamefont {Windey}}, \bibinfo {author} {\bibfnamefont
  {F.}~\bibnamefont {Tebbenjohanns}}, \ and\ \bibinfo {author} {\bibfnamefont
  {L.}~\bibnamefont {Novotny}},\ }\href@noop {} {\bibfield  {journal} {\bibinfo
   {journal} {Phys. Rev. Lett.}\ }\textbf {\bibinfo {volume} {121}},\ \bibinfo
  {pages} {033602} (\bibinfo {year} {2018})}\BibitemShut {NoStop}%
\bibitem [{\citenamefont {Ahn}\ \emph {et~al.}(2020)\citenamefont {Ahn},
  \citenamefont {Xu}, \citenamefont {Bang}, \citenamefont {Ju}, \citenamefont
  {Gao},\ and\ \citenamefont {Li}}]{ahn2020ultrasensitive}%
  \BibitemOpen
  \bibfield  {author} {\bibinfo {author} {\bibfnamefont {J.}~\bibnamefont
  {Ahn}}, \bibinfo {author} {\bibfnamefont {Z.}~\bibnamefont {Xu}}, \bibinfo
  {author} {\bibfnamefont {J.}~\bibnamefont {Bang}}, \bibinfo {author}
  {\bibfnamefont {P.}~\bibnamefont {Ju}}, \bibinfo {author} {\bibfnamefont
  {X.}~\bibnamefont {Gao}}, \ and\ \bibinfo {author} {\bibfnamefont
  {T.}~\bibnamefont {Li}},\ }\href@noop {} {\bibfield  {journal} {\bibinfo
  {journal} {Nat. Nanotechnol.}\ }\textbf {\bibinfo {volume} {15}},\ \bibinfo
  {pages} {89} (\bibinfo {year} {2020})}\BibitemShut {NoStop}%
\bibitem [{\citenamefont {Jin}\ \emph {et~al.}(2021)\citenamefont {Jin},
  \citenamefont {Yan}, \citenamefont {Rahman}, \citenamefont {Li},
  \citenamefont {Yu},\ and\ \citenamefont {Zhang}}]{jin20216}%
  \BibitemOpen
  \bibfield  {author} {\bibinfo {author} {\bibfnamefont {Y.}~\bibnamefont
  {Jin}}, \bibinfo {author} {\bibfnamefont {J.}~\bibnamefont {Yan}}, \bibinfo
  {author} {\bibfnamefont {S.~J.}\ \bibnamefont {Rahman}}, \bibinfo {author}
  {\bibfnamefont {J.}~\bibnamefont {Li}}, \bibinfo {author} {\bibfnamefont
  {X.}~\bibnamefont {Yu}}, \ and\ \bibinfo {author} {\bibfnamefont
  {J.}~\bibnamefont {Zhang}},\ }\href@noop {} {\bibfield  {journal} {\bibinfo
  {journal} {Photonics Res.}\ }\textbf {\bibinfo {volume} {9}},\ \bibinfo
  {pages} {1344} (\bibinfo {year} {2021})}\BibitemShut {NoStop}%
\bibitem [{\citenamefont {Manjavacas}\ and\ \citenamefont
  {de~Abajo}(2010)}]{manjavacas2010vacuum}%
  \BibitemOpen
  \bibfield  {author} {\bibinfo {author} {\bibfnamefont {A.}~\bibnamefont
  {Manjavacas}}\ and\ \bibinfo {author} {\bibfnamefont {F.~J.~G.}\ \bibnamefont
  {de~Abajo}},\ }\href@noop {} {\bibfield  {journal} {\bibinfo  {journal}
  {Phys. Rev. Lett.}\ }\textbf {\bibinfo {volume} {105}},\ \bibinfo {pages}
  {113601} (\bibinfo {year} {2010})}\BibitemShut {NoStop}%
\bibitem [{\citenamefont {Zhao}\ \emph {et~al.}(2012)\citenamefont {Zhao},
  \citenamefont {Manjavacas}, \citenamefont {de~Abajo},\ and\ \citenamefont
  {Pendry}}]{zhao2012rotational}%
  \BibitemOpen
  \bibfield  {author} {\bibinfo {author} {\bibfnamefont {R.~K.}\ \bibnamefont
  {Zhao}}, \bibinfo {author} {\bibfnamefont {A.}~\bibnamefont {Manjavacas}},
  \bibinfo {author} {\bibfnamefont {F.~J.~G.}\ \bibnamefont {de~Abajo}}, \ and\
  \bibinfo {author} {\bibfnamefont {J.~B.}\ \bibnamefont {Pendry}},\
  }\href@noop {} {\bibfield  {journal} {\bibinfo  {journal} {Phys. Rev. Lett.}\
  }\textbf {\bibinfo {volume} {109}},\ \bibinfo {pages} {123604} (\bibinfo
  {year} {2012})}\BibitemShut {NoStop}%
\bibitem [{\citenamefont {Stickler}\ \emph {et~al.}(2018)\citenamefont
  {Stickler}, \citenamefont {Schrinski},\ and\ \citenamefont
  {Hornberger}}]{stickler2018rotational}%
  \BibitemOpen
  \bibfield  {author} {\bibinfo {author} {\bibfnamefont {B.~A.}\ \bibnamefont
  {Stickler}}, \bibinfo {author} {\bibfnamefont {B.}~\bibnamefont {Schrinski}},
  \ and\ \bibinfo {author} {\bibfnamefont {K.}~\bibnamefont {Hornberger}},\
  }\href@noop {} {\bibfield  {journal} {\bibinfo  {journal} {Phys. Rev. Lett.}\
  }\textbf {\bibinfo {volume} {121}},\ \bibinfo {pages} {040401} (\bibinfo
  {year} {2018})}\BibitemShut {NoStop}%
\bibitem [{\citenamefont {Stickler}\ \emph {et~al.}(2021)\citenamefont
  {Stickler}, \citenamefont {Hornberger},\ and\ \citenamefont
  {Kim}}]{stickler2021quantum}%
  \BibitemOpen
  \bibfield  {author} {\bibinfo {author} {\bibfnamefont {B.~A.}\ \bibnamefont
  {Stickler}}, \bibinfo {author} {\bibfnamefont {K.}~\bibnamefont
  {Hornberger}}, \ and\ \bibinfo {author} {\bibfnamefont {M.}~\bibnamefont
  {Kim}},\ }\href@noop {} {\bibfield  {journal} {\bibinfo  {journal} {Nat. Rev.
  Phys.}\ }\textbf {\bibinfo {volume} {3}},\ \bibinfo {pages} {589} (\bibinfo
  {year} {2021})}\BibitemShut {NoStop}%
\bibitem [{\citenamefont {Svak}\ \emph
  {et~al.}(2018{\natexlab{a}})\citenamefont {Svak}, \citenamefont
  {Brzobohat{\`y}}, \citenamefont {{\v{S}}iler}, \citenamefont {J{\'a}kl},
  \citenamefont {Ka{\v{n}}ka}, \citenamefont {Zem{\'a}nek},\ and\ \citenamefont
  {Simpson}}]{Svak2018Trans}%
  \BibitemOpen
  \bibfield  {author} {\bibinfo {author} {\bibfnamefont {V.}~\bibnamefont
  {Svak}}, \bibinfo {author} {\bibfnamefont {O.}~\bibnamefont
  {Brzobohat{\`y}}}, \bibinfo {author} {\bibfnamefont {M.}~\bibnamefont
  {{\v{S}}iler}}, \bibinfo {author} {\bibfnamefont {P.}~\bibnamefont
  {J{\'a}kl}}, \bibinfo {author} {\bibfnamefont {J.}~\bibnamefont
  {Ka{\v{n}}ka}}, \bibinfo {author} {\bibfnamefont {P.}~\bibnamefont
  {Zem{\'a}nek}}, \ and\ \bibinfo {author} {\bibfnamefont {S.}~\bibnamefont
  {Simpson}},\ }\href@noop {} {\bibfield  {journal} {\bibinfo  {journal} {Nat.
  Commun.}\ }\textbf {\bibinfo {volume} {9}},\ \bibinfo {pages} {5453}
  (\bibinfo {year} {2018}{\natexlab{a}})}\BibitemShut {NoStop}%
\bibitem [{\citenamefont {Arita}\ \emph {et~al.}(2020)\citenamefont {Arita},
  \citenamefont {Simpson}, \citenamefont {Zem{\'a}nek},\ and\ \citenamefont
  {Dholakia}}]{arita2020coherent}%
  \BibitemOpen
  \bibfield  {author} {\bibinfo {author} {\bibfnamefont {Y.}~\bibnamefont
  {Arita}}, \bibinfo {author} {\bibfnamefont {S.~H.}\ \bibnamefont {Simpson}},
  \bibinfo {author} {\bibfnamefont {P.}~\bibnamefont {Zem{\'a}nek}}, \ and\
  \bibinfo {author} {\bibfnamefont {K.}~\bibnamefont {Dholakia}},\ }\href@noop
  {} {\bibfield  {journal} {\bibinfo  {journal} {Sci. Adv.}\ }\textbf {\bibinfo
  {volume} {6}},\ \bibinfo {pages} {eaaz9858} (\bibinfo {year}
  {2020})}\BibitemShut {NoStop}%
\bibitem [{\citenamefont {Schuck}\ \emph {et~al.}(2018)\citenamefont {Schuck},
  \citenamefont {Steinert}, \citenamefont {Nussbaumer},\ and\ \citenamefont
  {Kolar}}]{schuck2018ultrafast}%
  \BibitemOpen
  \bibfield  {author} {\bibinfo {author} {\bibfnamefont {M.}~\bibnamefont
  {Schuck}}, \bibinfo {author} {\bibfnamefont {D.}~\bibnamefont {Steinert}},
  \bibinfo {author} {\bibfnamefont {T.}~\bibnamefont {Nussbaumer}}, \ and\
  \bibinfo {author} {\bibfnamefont {J.~W.}\ \bibnamefont {Kolar}},\ }\href@noop
  {} {\bibfield  {journal} {\bibinfo  {journal} {Sci. Adv.}\ }\textbf {\bibinfo
  {volume} {4}},\ \bibinfo {pages} {e1701519} (\bibinfo {year}
  {2018})}\BibitemShut {NoStop}%
\bibitem [{\citenamefont {Gonzalez-Ballestero}\ \emph
  {et~al.}(2021)\citenamefont {Gonzalez-Ballestero}, \citenamefont
  {Aspelmeyer}, \citenamefont {Novotny}, \citenamefont {Quidant},\ and\
  \citenamefont {Romero-Isart}}]{gonzalez2021levitodynamics}%
  \BibitemOpen
  \bibfield  {author} {\bibinfo {author} {\bibfnamefont {C.}~\bibnamefont
  {Gonzalez-Ballestero}}, \bibinfo {author} {\bibfnamefont {M.}~\bibnamefont
  {Aspelmeyer}}, \bibinfo {author} {\bibfnamefont {L.}~\bibnamefont {Novotny}},
  \bibinfo {author} {\bibfnamefont {R.}~\bibnamefont {Quidant}}, \ and\
  \bibinfo {author} {\bibfnamefont {O.}~\bibnamefont {Romero-Isart}},\
  }\href@noop {} {\bibfield  {journal} {\bibinfo  {journal} {Science}\ }\textbf
  {\bibinfo {volume} {374}},\ \bibinfo {pages} {eabg3027} (\bibinfo {year}
  {2021})}\BibitemShut {NoStop}%
\bibitem [{\citenamefont {Walter}\ \emph {et~al.}(2015)\citenamefont {Walter},
  \citenamefont {Nunnenkamp},\ and\ \citenamefont
  {Bruder}}]{walter2015quantum}%
  \BibitemOpen
  \bibfield  {author} {\bibinfo {author} {\bibfnamefont {S.}~\bibnamefont
  {Walter}}, \bibinfo {author} {\bibfnamefont {A.}~\bibnamefont {Nunnenkamp}},
  \ and\ \bibinfo {author} {\bibfnamefont {C.}~\bibnamefont {Bruder}},\
  }\href@noop {} {\bibfield  {journal} {\bibinfo  {journal} {Ann. Phys.}\
  }\textbf {\bibinfo {volume} {527}},\ \bibinfo {pages} {131} (\bibinfo {year}
  {2015})}\BibitemShut {NoStop}%
\bibitem [{\citenamefont {Roulet}\ and\ \citenamefont
  {Bruder}(2018)}]{roulet2018quantum}%
  \BibitemOpen
  \bibfield  {author} {\bibinfo {author} {\bibfnamefont {A.}~\bibnamefont
  {Roulet}}\ and\ \bibinfo {author} {\bibfnamefont {C.}~\bibnamefont
  {Bruder}},\ }\href@noop {} {\bibfield  {journal} {\bibinfo  {journal} {Phys.
  Rev. Lett.}\ }\textbf {\bibinfo {volume} {121}},\ \bibinfo {pages} {063601}
  (\bibinfo {year} {2018})}\BibitemShut {NoStop}%
\bibitem [{\citenamefont {Kato}\ \emph {et~al.}(2019)\citenamefont {Kato},
  \citenamefont {Yamamoto},\ and\ \citenamefont
  {Nakao}}]{kato2019semiclassical}%
  \BibitemOpen
  \bibfield  {author} {\bibinfo {author} {\bibfnamefont {Y.}~\bibnamefont
  {Kato}}, \bibinfo {author} {\bibfnamefont {N.}~\bibnamefont {Yamamoto}}, \
  and\ \bibinfo {author} {\bibfnamefont {H.}~\bibnamefont {Nakao}},\
  }\href@noop {} {\bibfield  {journal} {\bibinfo  {journal} {Phys. Rev. Res.}\
  }\textbf {\bibinfo {volume} {1}},\ \bibinfo {pages} {033012} (\bibinfo {year}
  {2019})}\BibitemShut {NoStop}%
\bibitem [{\citenamefont {W{\"a}chtler}\ \emph {et~al.}(2019)\citenamefont
  {W{\"a}chtler}, \citenamefont {Strasberg}, \citenamefont {Klapp},
  \citenamefont {Schaller},\ and\ \citenamefont
  {Jarzynski}}]{wachtler2019stochastic}%
  \BibitemOpen
  \bibfield  {author} {\bibinfo {author} {\bibfnamefont {C.~W.}\ \bibnamefont
  {W{\"a}chtler}}, \bibinfo {author} {\bibfnamefont {P.}~\bibnamefont
  {Strasberg}}, \bibinfo {author} {\bibfnamefont {S.~H.}\ \bibnamefont
  {Klapp}}, \bibinfo {author} {\bibfnamefont {G.}~\bibnamefont {Schaller}}, \
  and\ \bibinfo {author} {\bibfnamefont {C.}~\bibnamefont {Jarzynski}},\
  }\href@noop {} {\bibfield  {journal} {\bibinfo  {journal} {New J. Phys.}\
  }\textbf {\bibinfo {volume} {21}},\ \bibinfo {pages} {073009} (\bibinfo
  {year} {2019})}\BibitemShut {NoStop}%
\bibitem [{\citenamefont {Bekshaev}\ \emph {et~al.}(2011)\citenamefont
  {Bekshaev}, \citenamefont {Bliokh},\ and\ \citenamefont
  {Soskin}}]{bekshaev2011internal}%
  \BibitemOpen
  \bibfield  {author} {\bibinfo {author} {\bibfnamefont {A.}~\bibnamefont
  {Bekshaev}}, \bibinfo {author} {\bibfnamefont {K.~Y.}\ \bibnamefont
  {Bliokh}}, \ and\ \bibinfo {author} {\bibfnamefont {M.}~\bibnamefont
  {Soskin}},\ }\href@noop {} {\bibfield  {journal} {\bibinfo  {journal} {J.
  Opt.}\ }\textbf {\bibinfo {volume} {13}},\ \bibinfo {pages} {053001}
  (\bibinfo {year} {2011})}\BibitemShut {NoStop}%
\bibitem [{\citenamefont {Simpson}\ \emph {et~al.}(2021)\citenamefont
  {Simpson}, \citenamefont {Arita}, \citenamefont {Dholakia},\ and\
  \citenamefont {Zem{\'a}nek}}]{simpson2021stochastic}%
  \BibitemOpen
  \bibfield  {author} {\bibinfo {author} {\bibfnamefont {S.~H.}\ \bibnamefont
  {Simpson}}, \bibinfo {author} {\bibfnamefont {Y.}~\bibnamefont {Arita}},
  \bibinfo {author} {\bibfnamefont {K.}~\bibnamefont {Dholakia}}, \ and\
  \bibinfo {author} {\bibfnamefont {P.}~\bibnamefont {Zem{\'a}nek}},\
  }\href@noop {} {\bibfield  {journal} {\bibinfo  {journal} {Phys. Rev. A}\
  }\textbf {\bibinfo {volume} {104}},\ \bibinfo {pages} {043518} (\bibinfo
  {year} {2021})}\BibitemShut {NoStop}%
\bibitem [{\citenamefont {Jones}\ \emph {et~al.}(2009)\citenamefont {Jones},
  \citenamefont {Palmisano}, \citenamefont {Bonaccorso}, \citenamefont
  {Gucciardi}, \citenamefont {Calogero}, \citenamefont {Ferrari},\ and\
  \citenamefont {Marago}}]{jones2009rotation}%
  \BibitemOpen
  \bibfield  {author} {\bibinfo {author} {\bibfnamefont {P.}~\bibnamefont
  {Jones}}, \bibinfo {author} {\bibfnamefont {F.}~\bibnamefont {Palmisano}},
  \bibinfo {author} {\bibfnamefont {F.}~\bibnamefont {Bonaccorso}}, \bibinfo
  {author} {\bibfnamefont {P.}~\bibnamefont {Gucciardi}}, \bibinfo {author}
  {\bibfnamefont {G.}~\bibnamefont {Calogero}}, \bibinfo {author}
  {\bibfnamefont {A.}~\bibnamefont {Ferrari}}, \ and\ \bibinfo {author}
  {\bibfnamefont {O.}~\bibnamefont {Marago}},\ }\href@noop {} {\bibfield
  {journal} {\bibinfo  {journal} {ACS Nano}\ }\textbf {\bibinfo {volume} {3}},\
  \bibinfo {pages} {3077} (\bibinfo {year} {2009})}\BibitemShut {NoStop}%
\bibitem [{\citenamefont {Svak}\ \emph
  {et~al.}(2018{\natexlab{b}})\citenamefont {Svak}, \citenamefont
  {Brzobohat{\`y}}, \citenamefont {{\v{S}}iler}, \citenamefont {J{\'a}kl},
  \citenamefont {Ka{\v{n}}ka}, \citenamefont {Zem{\'a}nek},\ and\ \citenamefont
  {Simpson}}]{svak2018transverse}%
  \BibitemOpen
  \bibfield  {author} {\bibinfo {author} {\bibfnamefont {V.}~\bibnamefont
  {Svak}}, \bibinfo {author} {\bibfnamefont {O.}~\bibnamefont
  {Brzobohat{\`y}}}, \bibinfo {author} {\bibfnamefont {M.}~\bibnamefont
  {{\v{S}}iler}}, \bibinfo {author} {\bibfnamefont {P.}~\bibnamefont
  {J{\'a}kl}}, \bibinfo {author} {\bibfnamefont {J.}~\bibnamefont
  {Ka{\v{n}}ka}}, \bibinfo {author} {\bibfnamefont {P.}~\bibnamefont
  {Zem{\'a}nek}}, \ and\ \bibinfo {author} {\bibfnamefont {S.}~\bibnamefont
  {Simpson}},\ }\href@noop {} {\bibfield  {journal} {\bibinfo  {journal} {Nat.
  Commun.}\ }\textbf {\bibinfo {volume} {9}},\ \bibinfo {pages} {1} (\bibinfo
  {year} {2018}{\natexlab{b}})}\BibitemShut {NoStop}%
\bibitem [{\citenamefont {Gieseler}\ \emph {et~al.}(2015)\citenamefont
  {Gieseler}, \citenamefont {Novotny}, \citenamefont {Moritz},\ and\
  \citenamefont {Dellago}}]{gieseler2015non}%
  \BibitemOpen
  \bibfield  {author} {\bibinfo {author} {\bibfnamefont {J.}~\bibnamefont
  {Gieseler}}, \bibinfo {author} {\bibfnamefont {L.}~\bibnamefont {Novotny}},
  \bibinfo {author} {\bibfnamefont {C.}~\bibnamefont {Moritz}}, \ and\ \bibinfo
  {author} {\bibfnamefont {C.}~\bibnamefont {Dellago}},\ }\href@noop {}
  {\bibfield  {journal} {\bibinfo  {journal} {New J. Phys.}\ }\textbf {\bibinfo
  {volume} {17}},\ \bibinfo {pages} {045011} (\bibinfo {year}
  {2015})}\BibitemShut {NoStop}%
\bibitem [{\citenamefont {Van~Kampen}(1992)}]{van1992stochastic}%
  \BibitemOpen
  \bibfield  {author} {\bibinfo {author} {\bibfnamefont {N.~G.}\ \bibnamefont
  {Van~Kampen}},\ }\href@noop {} {\emph {\bibinfo {title} {Stochastic processes
  in physics and chemistry}}},\ Vol.~\bibinfo {volume} {1}\ (\bibinfo
  {publisher} {Elsevier},\ \bibinfo {year} {1992})\BibitemShut {NoStop}%
\bibitem [{\citenamefont {Pikovsky}\ \emph {et~al.}(2002)\citenamefont
  {Pikovsky}, \citenamefont {Rosenblum},\ and\ \citenamefont
  {Kurths}}]{pikovsky2002synchronization}%
  \BibitemOpen
  \bibfield  {author} {\bibinfo {author} {\bibfnamefont {A.}~\bibnamefont
  {Pikovsky}}, \bibinfo {author} {\bibfnamefont {M.}~\bibnamefont {Rosenblum}},
  \ and\ \bibinfo {author} {\bibfnamefont {J.}~\bibnamefont {Kurths}},\
  }\href@noop {} {\enquote {\bibinfo {title} {Synchronization: a universal
  concept in nonlinear science},}\ } (\bibinfo {year} {2002})\BibitemShut
  {NoStop}%
\bibitem [{\citenamefont {Pikovsky}\ \emph {et~al.}(2000)\citenamefont
  {Pikovsky}, \citenamefont {Rosenblum},\ and\ \citenamefont
  {Kurths}}]{pikovsky2000phase}%
  \BibitemOpen
  \bibfield  {author} {\bibinfo {author} {\bibfnamefont {A.}~\bibnamefont
  {Pikovsky}}, \bibinfo {author} {\bibfnamefont {M.}~\bibnamefont {Rosenblum}},
  \ and\ \bibinfo {author} {\bibfnamefont {J.}~\bibnamefont {Kurths}},\
  }\href@noop {} {\bibfield  {journal} {\bibinfo  {journal} {Int. J. Bifurc.
  Chaos Appl. Sci. Eng.}\ }\textbf {\bibinfo {volume} {10}},\ \bibinfo {pages}
  {2291} (\bibinfo {year} {2000})}\BibitemShut {NoStop}%
\bibitem [{\citenamefont {Magallanes}\ and\ \citenamefont
  {Brasselet}(2018)}]{magallanes2018macroscopic}%
  \BibitemOpen
  \bibfield  {author} {\bibinfo {author} {\bibfnamefont {H.}~\bibnamefont
  {Magallanes}}\ and\ \bibinfo {author} {\bibfnamefont {E.}~\bibnamefont
  {Brasselet}},\ }\href@noop {} {\bibfield  {journal} {\bibinfo  {journal}
  {Nat. Photon.}\ }\textbf {\bibinfo {volume} {12}},\ \bibinfo {pages} {461}
  (\bibinfo {year} {2018})}\BibitemShut {NoStop}%
\bibitem [{\citenamefont {Brzobohat{\`y}}\ \emph {et~al.}(2013)\citenamefont
  {Brzobohat{\`y}}, \citenamefont {Kar{\'a}sek}, \citenamefont {{\v{S}}iler},
  \citenamefont {Chv{\'a}tal}, \citenamefont {{\v{C}}i{\v{z}}m{\'a}r},\ and\
  \citenamefont {Zem{\'a}nek}}]{brzobohaty2013experimental}%
  \BibitemOpen
  \bibfield  {author} {\bibinfo {author} {\bibfnamefont {O.}~\bibnamefont
  {Brzobohat{\`y}}}, \bibinfo {author} {\bibfnamefont {V.}~\bibnamefont
  {Kar{\'a}sek}}, \bibinfo {author} {\bibfnamefont {M.}~\bibnamefont
  {{\v{S}}iler}}, \bibinfo {author} {\bibfnamefont {L.}~\bibnamefont
  {Chv{\'a}tal}}, \bibinfo {author} {\bibfnamefont {T.}~\bibnamefont
  {{\v{C}}i{\v{z}}m{\'a}r}}, \ and\ \bibinfo {author} {\bibfnamefont
  {P.}~\bibnamefont {Zem{\'a}nek}},\ }\href@noop {} {\bibfield  {journal}
  {\bibinfo  {journal} {Nat. Photon.}\ }\textbf {\bibinfo {volume} {7}},\
  \bibinfo {pages} {123} (\bibinfo {year} {2013})}\BibitemShut {NoStop}%
\bibitem [{\citenamefont {Han}\ \emph {et~al.}(2018)\citenamefont {Han},
  \citenamefont {Parker}, \citenamefont {Yifat}, \citenamefont {Peterson},
  \citenamefont {Gray}, \citenamefont {Scherer},\ and\ \citenamefont
  {Yan}}]{han2018crossover}%
  \BibitemOpen
  \bibfield  {author} {\bibinfo {author} {\bibfnamefont {F.}~\bibnamefont
  {Han}}, \bibinfo {author} {\bibfnamefont {J.~A.}\ \bibnamefont {Parker}},
  \bibinfo {author} {\bibfnamefont {Y.}~\bibnamefont {Yifat}}, \bibinfo
  {author} {\bibfnamefont {C.}~\bibnamefont {Peterson}}, \bibinfo {author}
  {\bibfnamefont {S.~K.}\ \bibnamefont {Gray}}, \bibinfo {author}
  {\bibfnamefont {N.~F.}\ \bibnamefont {Scherer}}, \ and\ \bibinfo {author}
  {\bibfnamefont {Z.}~\bibnamefont {Yan}},\ }\href@noop {} {\bibfield
  {journal} {\bibinfo  {journal} {Nat. Commun.}\ }\textbf {\bibinfo {volume}
  {9}},\ \bibinfo {pages} {1} (\bibinfo {year} {2018})}\BibitemShut {NoStop}%
\bibitem [{\citenamefont {Simpson}\ and\ \citenamefont
  {Hanna}(2007)}]{simpson2007optical}%
  \BibitemOpen
  \bibfield  {author} {\bibinfo {author} {\bibfnamefont {S.~H.}\ \bibnamefont
  {Simpson}}\ and\ \bibinfo {author} {\bibfnamefont {S.}~\bibnamefont
  {Hanna}},\ }\href@noop {} {\bibfield  {journal} {\bibinfo  {journal} {J. Opt.
  Soc. Am. A}\ }\textbf {\bibinfo {volume} {24}},\ \bibinfo {pages} {430}
  (\bibinfo {year} {2007})}\BibitemShut {NoStop}%
\bibitem [{\citenamefont {Diniz}\ \emph {et~al.}(2019)\citenamefont {Diniz},
  \citenamefont {Dutra}, \citenamefont {Pires}, \citenamefont {Viana},
  \citenamefont {Nussenzveig},\ and\ \citenamefont {Neto}}]{diniz2019negative}%
  \BibitemOpen
  \bibfield  {author} {\bibinfo {author} {\bibfnamefont {K.}~\bibnamefont
  {Diniz}}, \bibinfo {author} {\bibfnamefont {R.}~\bibnamefont {Dutra}},
  \bibinfo {author} {\bibfnamefont {L.}~\bibnamefont {Pires}}, \bibinfo
  {author} {\bibfnamefont {N.}~\bibnamefont {Viana}}, \bibinfo {author}
  {\bibfnamefont {H.}~\bibnamefont {Nussenzveig}}, \ and\ \bibinfo {author}
  {\bibfnamefont {P.~M.}\ \bibnamefont {Neto}},\ }\href@noop {} {\bibfield
  {journal} {\bibinfo  {journal} {Opt. Express}\ }\textbf {\bibinfo {volume}
  {27}},\ \bibinfo {pages} {5905} (\bibinfo {year} {2019})}\BibitemShut
  {NoStop}%
\bibitem [{\citenamefont {Bliokh}\ \emph {et~al.}(2014)\citenamefont {Bliokh},
  \citenamefont {Bekshaev},\ and\ \citenamefont
  {Nori}}]{bliokh2014extraordinary}%
  \BibitemOpen
  \bibfield  {author} {\bibinfo {author} {\bibfnamefont {K.~Y.}\ \bibnamefont
  {Bliokh}}, \bibinfo {author} {\bibfnamefont {A.~Y.}\ \bibnamefont
  {Bekshaev}}, \ and\ \bibinfo {author} {\bibfnamefont {F.}~\bibnamefont
  {Nori}},\ }\href@noop {} {\bibfield  {journal} {\bibinfo  {journal} {Nat.
  Commun.}\ }\textbf {\bibinfo {volume} {5}},\ \bibinfo {pages} {1} (\bibinfo
  {year} {2014})}\BibitemShut {NoStop}%
\bibitem [{\citenamefont {Antognozzi}\ \emph {et~al.}(2016)\citenamefont
  {Antognozzi}, \citenamefont {Bermingham}, \citenamefont {Harniman},
  \citenamefont {Simpson}, \citenamefont {Senior}, \citenamefont {Hayward},
  \citenamefont {Hoerber}, \citenamefont {Dennis}, \citenamefont {Bekshaev},
  \citenamefont {Bliokh} \emph {et~al.}}]{antognozzi2016direct}%
  \BibitemOpen
  \bibfield  {author} {\bibinfo {author} {\bibfnamefont {M.}~\bibnamefont
  {Antognozzi}}, \bibinfo {author} {\bibfnamefont {C.}~\bibnamefont
  {Bermingham}}, \bibinfo {author} {\bibfnamefont {R.}~\bibnamefont
  {Harniman}}, \bibinfo {author} {\bibfnamefont {S.}~\bibnamefont {Simpson}},
  \bibinfo {author} {\bibfnamefont {J.}~\bibnamefont {Senior}}, \bibinfo
  {author} {\bibfnamefont {R.}~\bibnamefont {Hayward}}, \bibinfo {author}
  {\bibfnamefont {H.}~\bibnamefont {Hoerber}}, \bibinfo {author} {\bibfnamefont
  {M.}~\bibnamefont {Dennis}}, \bibinfo {author} {\bibfnamefont
  {A.}~\bibnamefont {Bekshaev}}, \bibinfo {author} {\bibfnamefont
  {K.}~\bibnamefont {Bliokh}},  \emph {et~al.},\ }\href@noop {} {\bibfield
  {journal} {\bibinfo  {journal} {Nat. Phys.}\ }\textbf {\bibinfo {volume}
  {12}},\ \bibinfo {pages} {731} (\bibinfo {year} {2016})}\BibitemShut
  {NoStop}%
\bibitem [{\citenamefont {Amitai}\ \emph {et~al.}(2017)\citenamefont {Amitai},
  \citenamefont {L{\"o}rch}, \citenamefont {Nunnenkamp}, \citenamefont
  {Walter},\ and\ \citenamefont {Bruder}}]{amitai2017synchronization}%
  \BibitemOpen
  \bibfield  {author} {\bibinfo {author} {\bibfnamefont {E.}~\bibnamefont
  {Amitai}}, \bibinfo {author} {\bibfnamefont {N.}~\bibnamefont {L{\"o}rch}},
  \bibinfo {author} {\bibfnamefont {A.}~\bibnamefont {Nunnenkamp}}, \bibinfo
  {author} {\bibfnamefont {S.}~\bibnamefont {Walter}}, \ and\ \bibinfo {author}
  {\bibfnamefont {C.}~\bibnamefont {Bruder}},\ }\href@noop {} {\bibfield
  {journal} {\bibinfo  {journal} {Phys. Rev. A}\ }\textbf {\bibinfo {volume}
  {95}},\ \bibinfo {pages} {053858} (\bibinfo {year} {2017})}\BibitemShut
  {NoStop}%
\bibitem [{\citenamefont {Dieterich}\ \emph {et~al.}(2015)\citenamefont
  {Dieterich}, \citenamefont {Camunas-Soler}, \citenamefont
  {Ribezzi-Crivellari}, \citenamefont {Seifert},\ and\ \citenamefont
  {Ritort}}]{dieterich2015single}%
  \BibitemOpen
  \bibfield  {author} {\bibinfo {author} {\bibfnamefont {E.}~\bibnamefont
  {Dieterich}}, \bibinfo {author} {\bibfnamefont {J.}~\bibnamefont
  {Camunas-Soler}}, \bibinfo {author} {\bibfnamefont {M.}~\bibnamefont
  {Ribezzi-Crivellari}}, \bibinfo {author} {\bibfnamefont {U.}~\bibnamefont
  {Seifert}}, \ and\ \bibinfo {author} {\bibfnamefont {F.}~\bibnamefont
  {Ritort}},\ }\href@noop {} {\bibfield  {journal} {\bibinfo  {journal} {Nat.
  Phys.}\ }\textbf {\bibinfo {volume} {11}},\ \bibinfo {pages} {971} (\bibinfo
  {year} {2015})}\BibitemShut {NoStop}%
\bibitem [{\citenamefont {Casas-V{\'a}zquez}\ and\ \citenamefont
  {Jou}(2003)}]{casas2003temperature}%
  \BibitemOpen
  \bibfield  {author} {\bibinfo {author} {\bibfnamefont {J.}~\bibnamefont
  {Casas-V{\'a}zquez}}\ and\ \bibinfo {author} {\bibfnamefont {D.}~\bibnamefont
  {Jou}},\ }\href@noop {} {\bibfield  {journal} {\bibinfo  {journal} {Rep.
  Prog. Phys.}\ }\textbf {\bibinfo {volume} {66}},\ \bibinfo {pages} {1937}
  (\bibinfo {year} {2003})}\BibitemShut {NoStop}%
\bibitem [{\citenamefont {Aspelmeyer}\ \emph {et~al.}(2014)\citenamefont
  {Aspelmeyer}, \citenamefont {Kippenberg},\ and\ \citenamefont
  {Marquardt}}]{aspelmeyer2014cavity}%
  \BibitemOpen
  \bibfield  {author} {\bibinfo {author} {\bibfnamefont {M.}~\bibnamefont
  {Aspelmeyer}}, \bibinfo {author} {\bibfnamefont {T.~J.}\ \bibnamefont
  {Kippenberg}}, \ and\ \bibinfo {author} {\bibfnamefont {F.}~\bibnamefont
  {Marquardt}},\ }\href@noop {} {\bibfield  {journal} {\bibinfo  {journal}
  {Rev. Mod. Phys.}\ }\textbf {\bibinfo {volume} {86}},\ \bibinfo {pages}
  {1391} (\bibinfo {year} {2014})}\BibitemShut {NoStop}%
\bibitem [{\citenamefont {Witthaut}\ \emph {et~al.}(2017)\citenamefont
  {Witthaut}, \citenamefont {Wimberger}, \citenamefont {Burioni},\ and\
  \citenamefont {Timme}}]{witthaut2017classical}%
  \BibitemOpen
  \bibfield  {author} {\bibinfo {author} {\bibfnamefont {D.}~\bibnamefont
  {Witthaut}}, \bibinfo {author} {\bibfnamefont {S.}~\bibnamefont {Wimberger}},
  \bibinfo {author} {\bibfnamefont {R.}~\bibnamefont {Burioni}}, \ and\
  \bibinfo {author} {\bibfnamefont {M.}~\bibnamefont {Timme}},\ }\href@noop {}
  {\bibfield  {journal} {\bibinfo  {journal} {Nat. Commun.}\ }\textbf {\bibinfo
  {volume} {8}},\ \bibinfo {pages} {1} (\bibinfo {year} {2017})}\BibitemShut
  {NoStop}%
\bibitem [{\citenamefont {Neugebauer}\ \emph {et~al.}(2015)\citenamefont
  {Neugebauer}, \citenamefont {Bauer}, \citenamefont {Aiello},\ and\
  \citenamefont {Banzer}}]{neugebauer2015measuring}%
  \BibitemOpen
  \bibfield  {author} {\bibinfo {author} {\bibfnamefont {M.}~\bibnamefont
  {Neugebauer}}, \bibinfo {author} {\bibfnamefont {T.}~\bibnamefont {Bauer}},
  \bibinfo {author} {\bibfnamefont {A.}~\bibnamefont {Aiello}}, \ and\ \bibinfo
  {author} {\bibfnamefont {P.}~\bibnamefont {Banzer}},\ }\href@noop {}
  {\bibfield  {journal} {\bibinfo  {journal} {Phys. Rev. Lett.}\ }\textbf
  {\bibinfo {volume} {114}},\ \bibinfo {pages} {063901} (\bibinfo {year}
  {2015})}\BibitemShut {NoStop}%
\bibitem [{\citenamefont {Sun}\ \emph {et~al.}(2008)\citenamefont {Sun},
  \citenamefont {Lin},\ and\ \citenamefont {Gezelter}}]{sun2008langevin}%
  \BibitemOpen
  \bibfield  {author} {\bibinfo {author} {\bibfnamefont {X.}~\bibnamefont
  {Sun}}, \bibinfo {author} {\bibfnamefont {T.}~\bibnamefont {Lin}}, \ and\
  \bibinfo {author} {\bibfnamefont {J.~D.}\ \bibnamefont {Gezelter}},\
  }\href@noop {} {\bibfield  {journal} {\bibinfo  {journal} {J. Chem. Phys.}\
  }\textbf {\bibinfo {volume} {128}},\ \bibinfo {pages} {234107} (\bibinfo
  {year} {2008})}\BibitemShut {NoStop}%
\bibitem [{\citenamefont {Simpson}\ and\ \citenamefont
  {Hanna}(2009)}]{Simpson2009Opt}%
  \BibitemOpen
  \bibfield  {author} {\bibinfo {author} {\bibfnamefont {S.~H.}\ \bibnamefont
  {Simpson}}\ and\ \bibinfo {author} {\bibfnamefont {S.}~\bibnamefont
  {Hanna}},\ }\href@noop {} {\bibfield  {journal} {\bibinfo  {journal} {J. Opt.
  Soc. Am. A}\ }\textbf {\bibinfo {volume} {26}},\ \bibinfo {pages} {625}
  (\bibinfo {year} {2009})}\BibitemShut {NoStop}%
\bibitem [{\citenamefont {Coffey}\ and\ \citenamefont
  {Kalmykov}(2012)}]{coffey2012langevin}%
  \BibitemOpen
  \bibfield  {author} {\bibinfo {author} {\bibfnamefont {W.}~\bibnamefont
  {Coffey}}\ and\ \bibinfo {author} {\bibfnamefont {Y.~P.}\ \bibnamefont
  {Kalmykov}},\ }\href@noop {} {\emph {\bibinfo {title} {The Langevin equation:
  with applications to stochastic problems in physics, chemistry and electrical
  engineering}}},\ Vol.~\bibinfo {volume} {27}\ (\bibinfo  {publisher} {World
  Scientific},\ \bibinfo {year} {2012})\BibitemShut {NoStop}%
\bibitem [{\citenamefont {Sihvola}(1994)}]{sihvola1994dielectric}%
  \BibitemOpen
  \bibfield  {author} {\bibinfo {author} {\bibfnamefont {A.~H.}\ \bibnamefont
  {Sihvola}},\ }\href@noop {} {\bibfield  {journal} {\bibinfo  {journal} {Opt.
  Lett.}\ }\textbf {\bibinfo {volume} {19}},\ \bibinfo {pages} {430} (\bibinfo
  {year} {1994})}\BibitemShut {NoStop}%
\bibitem [{\citenamefont {Simpson}\ \emph {et~al.}(2016)\citenamefont
  {Simpson}, \citenamefont {Chv{\'a}tal},\ and\ \citenamefont
  {Zem{\'a}nek}}]{simpson2016synchronization}%
  \BibitemOpen
  \bibfield  {author} {\bibinfo {author} {\bibfnamefont {S.}~\bibnamefont
  {Simpson}}, \bibinfo {author} {\bibfnamefont {L.}~\bibnamefont
  {Chv{\'a}tal}}, \ and\ \bibinfo {author} {\bibfnamefont {P.}~\bibnamefont
  {Zem{\'a}nek}},\ }\href@noop {} {\bibfield  {journal} {\bibinfo  {journal}
  {Phys Rev. A}\ }\textbf {\bibinfo {volume} {93}},\ \bibinfo {pages} {023842}
  (\bibinfo {year} {2016})}\BibitemShut {NoStop}%
\bibitem [{\citenamefont {Chaumet}\ and\ \citenamefont
  {Nieto-Vesperinas}(2000)}]{chaumet2000time}%
  \BibitemOpen
  \bibfield  {author} {\bibinfo {author} {\bibfnamefont {P.~C.}\ \bibnamefont
  {Chaumet}}\ and\ \bibinfo {author} {\bibfnamefont {M.}~\bibnamefont
  {Nieto-Vesperinas}},\ }\href@noop {} {\bibfield  {journal} {\bibinfo
  {journal} {Opt. Lett.}\ }\textbf {\bibinfo {volume} {25}},\ \bibinfo {pages}
  {1065} (\bibinfo {year} {2000})}\BibitemShut {NoStop}%
\end{thebibliography}%


%

\clearpage

\setcounter{section}{0}
\setcounter{equation}{0}
\setcounter{figure}{0}
\setcounter{table}{0}
\setcounter{page}{1}
\makeatletter
\renewcommand{\thesection}{S\arabic{section}}
\renewcommand{\theequation}{S\arabic{equation}}
\renewcommand{\thefigure}{S\arabic{figure}}
\renewcommand{\thetable}{S\arabic{table}}

\widetext
\begin{center}
\textbf{
\large 
Cooling the optical-spin driven limit cycle oscillations of a levitated gyroscope: Supplementary information
}\\
\end{center}


\section{Anisotropic light scattering of vaterite}\label{sec:si_anisotropy}

\begin{figure*}[htb!]
\centering
\includegraphics[width = .5\textwidth, clip = true]{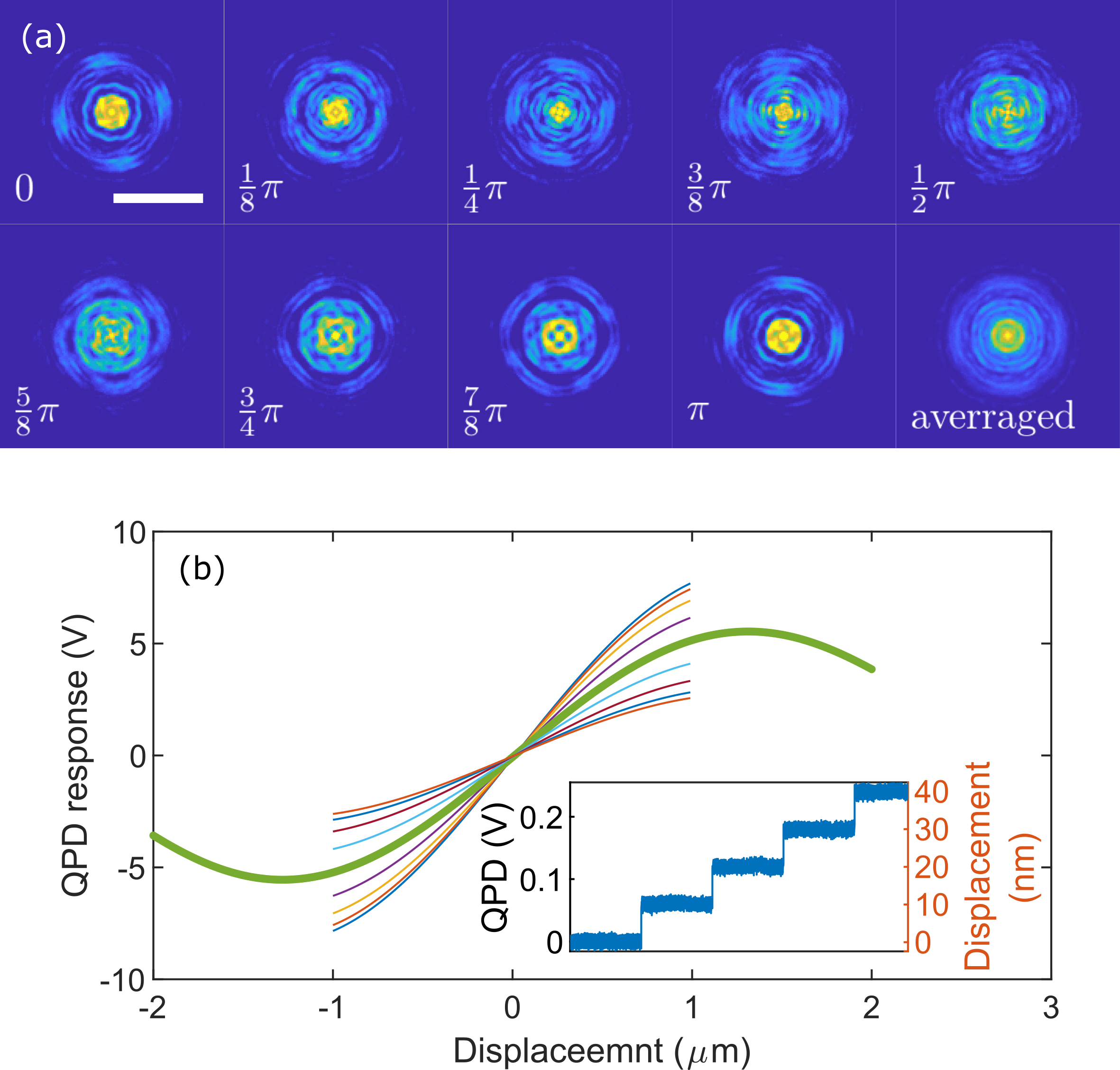}
\caption{
Anisotropic light scattering of vaterite. (a) Orientation dependent light scattering of vaterite in a circularly polarised trap and (b) corresponding QPD responses with respect to displacement, where the green curve represents a mean response of $5.59\,\mathrm{mV\,nm^{-1}}$ in its linear range. Inset shows the raw QPD signal with a noise of $12.2\,\mathrm{mV}$ ($2\sigma$) for steps of $10\,\mathrm{nm}$ displacements, indicating the position sensitivity of $2.0\,\mathrm{nm}\,(2\sigma)$. The scale bar in (a) indicates $5\,\upmu\mathrm{m}$.
}
\label{fig:vaterite_cp_back_scattering_}
\end{figure*}

\noindent When a birefringent uniaxial crystal, such as vaterite, is trapped in a circularly polarised beam, the optical axis of the crystal follows the rotating electric field, which is perpendicular to the beam propagation direction, causing the particle to rotate. Unlike a silica microsphere, light scattering by a birefringent microsphere is determined by its direction about the beam axis. Figure~\ref{fig:vaterite_cp_back_scattering_}(a) shows back-scattered light from a rotating vaterite microsphere trapped by circularly polarised light, which changes its intensity profile. Therefore care must be taken in the particle position detection using a QPD as the voltage response to its displacement relative to the beam is orientation-dependent (see Fig.~(\ref{fig:vaterite_cp_back_scattering_})b and the Methods section).

\section{Isotropic particle in a circularly polarised trap}\label{sec:si_isotropic}
\begin{figure*}[htb!]
\centering
\includegraphics[width = \textwidth, clip = true]{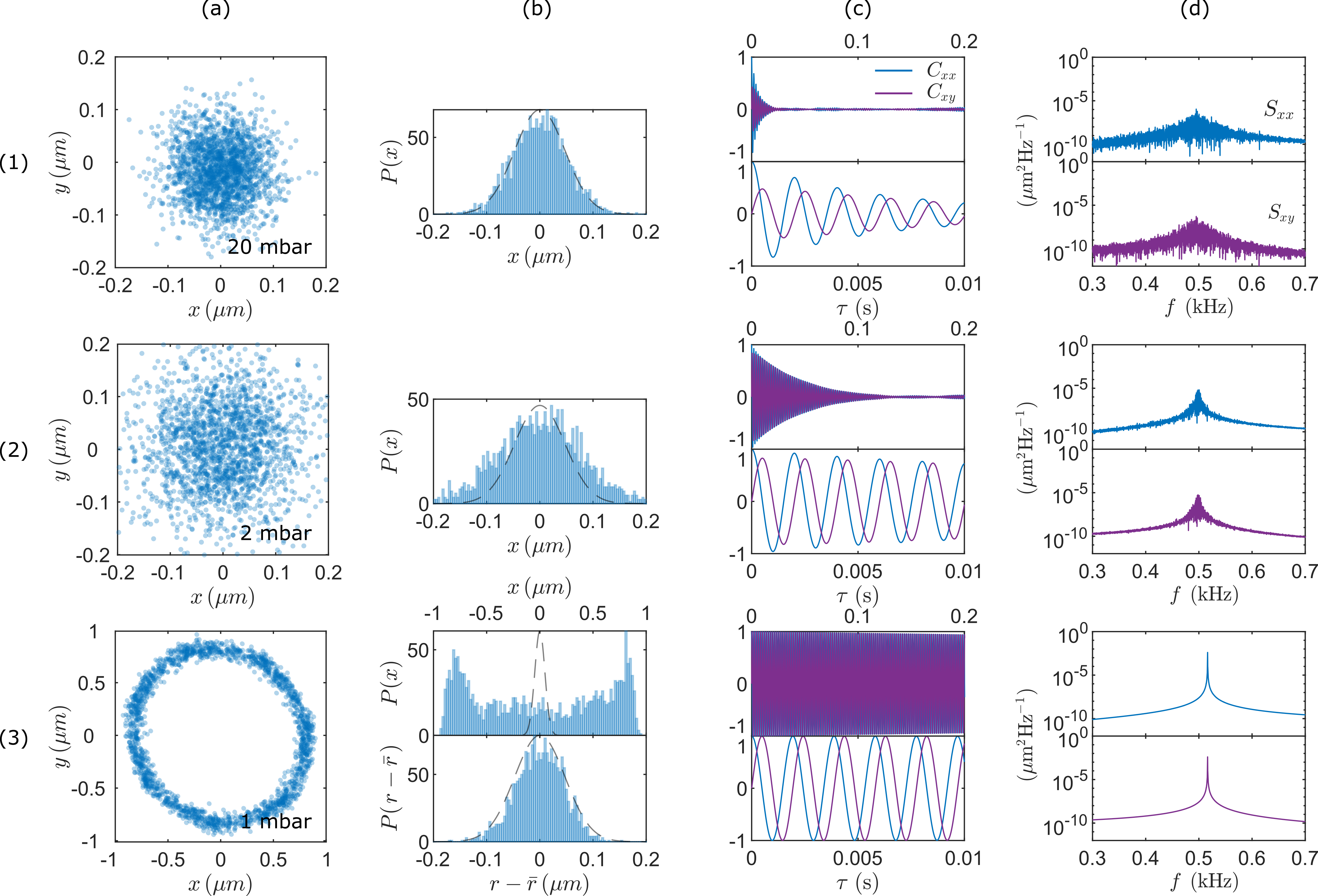}
\caption{
Experimentally measured dynamics of a silica microsphere trapped with a circularly polarised beam. (a) CoM position distributions in the $x-y$ plane (transverse to the beam axis, $z$) and (b) their histograms in terms of $x$ in Cartesian coordinates and the radial $r$ position in radial coordinates, where $\bar{r}$ indicates the mean radial position. The dashed lines show a Gaussian position distribution acquired at $20\,\mathrm{mbar}$. (c) Autocorrelation, $C_{xx}=\langle x(t)x(t+\tau) \rangle$ (blue) and cross correlation of $C_{xy}=\langle x(t) y(t+\tau) \rangle$ (purple), where their decay times $\tau_{D}$ are $3.7\,\mathrm{ms}$, $5.5\,\mathrm{ms}$ ($20\,\mathrm{mbar}$); $24\,\mathrm{ms}$, $26\,\mathrm{ms}$ ($2\,\mathrm{mbar}$); $52\,\mathrm{s}$, $53\,\mathrm{s}$ ($1\,\mathrm{mbar}$), respectively. (d) Power spectra of $C_{xx}=x(t)x(t+\tau)$ and $C_{xy}=x(t)y(t+\tau)$ showing the trap frequency at $f_x\approx f_y\sim0.5\,\mathrm{kHz}$. Rows (1) to (3) represent data at different gas pressures.
}
\label{fig:silica_cp_pmod_0_0_0_}
\end{figure*}

\noindent In a circularly polarised Gaussian trap in vacuum, isotropic spheres exhibit a range of nonequilibrium behaviour, which includes (i) biased stochastic motion; (ii) orbital motion; (iii) loss of particle, as the gas viscosity is decreased or the laser power is increased~\cite{Svak2018Trans}. These phenomena are associated with a linear component of optical momentum or transverse spin forces (TSFs) present in circularly polarised light~\cite{neugebauer2015measuring}. TSFs have been directly measured in evanescent waves~\cite{bliokh2014extraordinary,antognozzi2016direct} and in circularly polarised Gaussian traps~\cite{Svak2018Trans}. Supplementary Figure~\ref{fig:silica_cp_pmod_0_0_0_} shows the square root of the variance in the position of trapped silica microspheres (five samples with a radius of $2.5\,\upmu\mathrm{m}$) in a circularly polarised beam with a power of $15\,\mathrm{mW}$. The position variance increases with decreasing the gas pressure (or viscosity) towards $1-2\,\mathrm{mbar}$, where the trapped silica particles are ejected from the trap. This is a signature of the TSFs destabilising the trap~\cite{Svak2018Trans}. 

\section{Detailed simulations of the motion of a birefringent microsphere in a circularly polarized beam}\label{sec:si_detsims}
\begin{figure}[H]
\begin{tabular}{cc}
\resizebox{8cm}{!}{\includegraphics{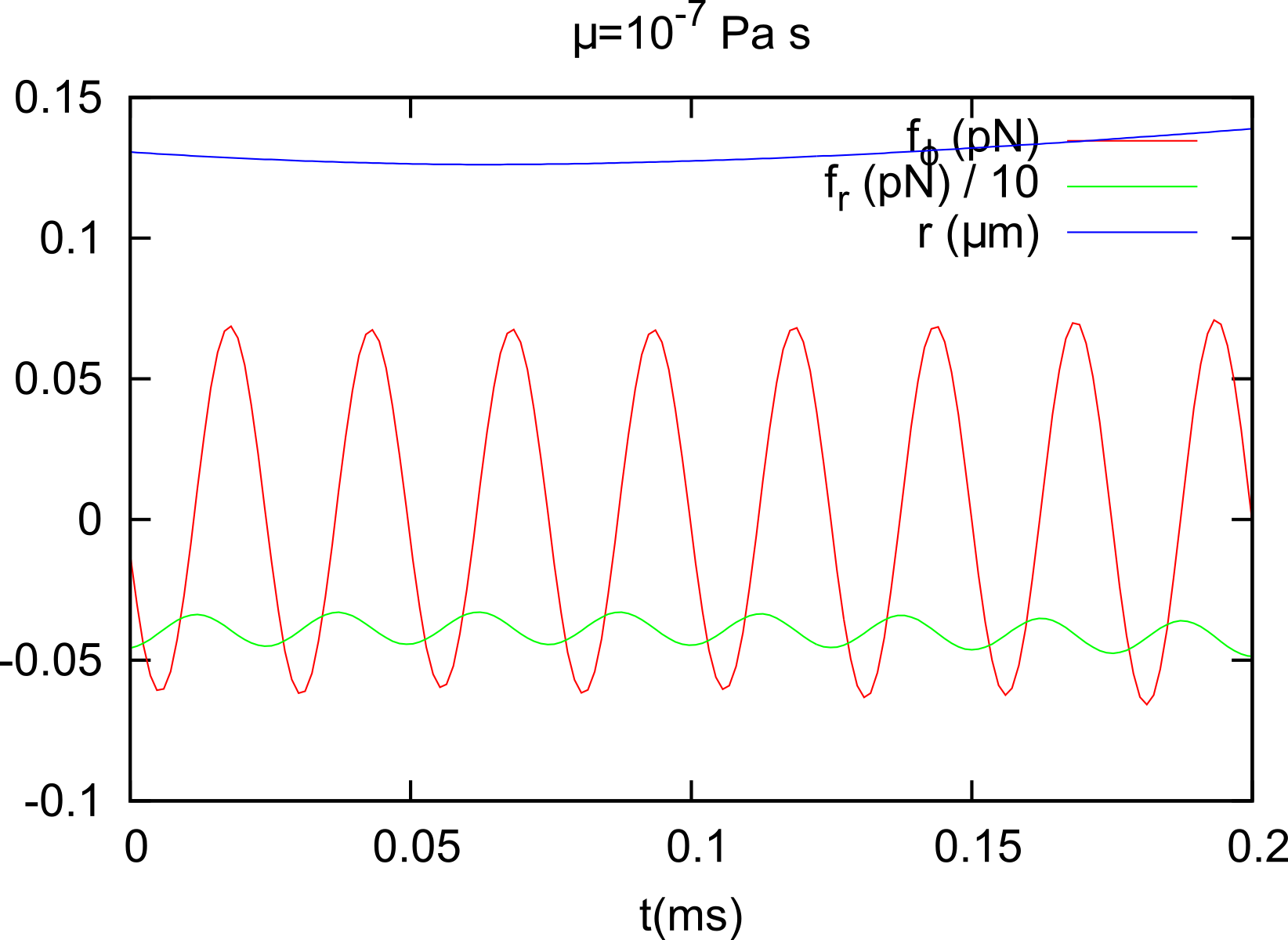}}  
\resizebox{8cm}{!}{\includegraphics{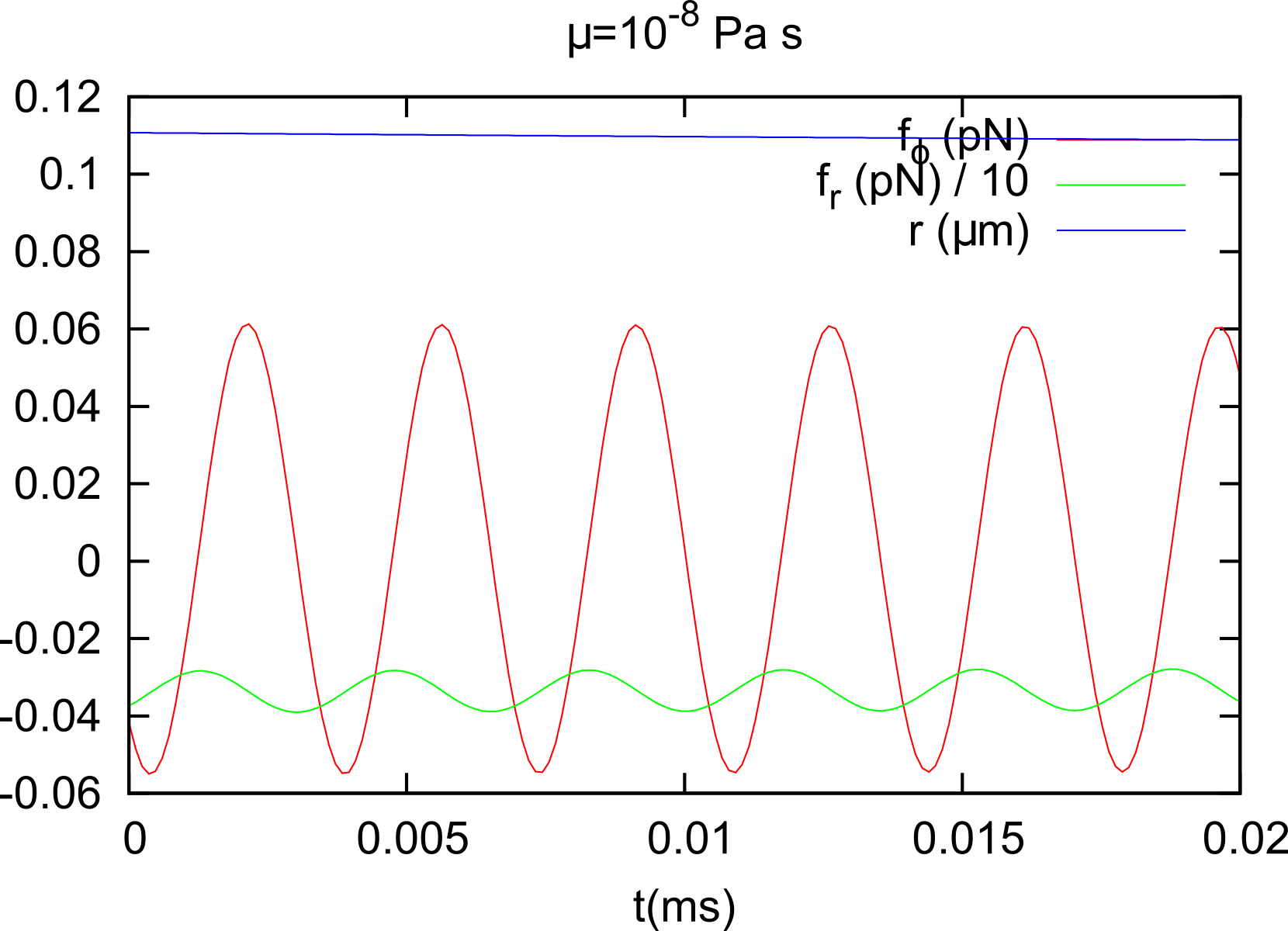}} 
\end{tabular}
\caption{\label{fig:sims_det2} Transverse forces and radial coordinate of spinning birefringent microsphere at two different viscosities, (a) $\mu=10^{-7}$Pas and (b) $\mu=10^{-8}$ Pas. The orientation dependence of the transverse forces results in oscillations as the particle spins}
\end{figure}

\noindent Direct simulations are performed in 3d by numerical integration of the following Langevin equation,
\begin{equation}\label{eq:SI_lang0}
\fvec^{opt}(\qvec)+\fvec^L(t) -mg \uz - \Ximat \dot \qvec = \Mmat \ddot \qvec.
\end{equation}
Where $\qvec$ are the coordinates of the centre mass $(x,y,z)$ and orientation of the rotationally symmetric particle. $\fvec^L(t)$ is the uncorrelated Langevin force, with amplitude fixed by the fluctuation-dissipation theorem and a mean value of zero.
\begin{eqnarray}\label{eq:flang}
\langle \fvec^L (t) \rangle &=& 0,\\
\langle \fvec^L(t) \otimes \fvec^L(t') \rangle &=& 2k_BT\Ximat\delta(t-t')
\end{eqnarray}
 $\Ximat$ is the hydrodynamic friction for a sphere of radius $a$ (with diagonal entries $\xi_t=6\pi\mu a$ for translations and $\xi_r=8\pi\mu a^3$ for rotations with viscosity $\mu$), $\Mmat$ the mass ($m$) and moment of inertia, ($I$). Viscosity and pressure are related according to,
\begin{equation}
\mu=\mu^{at}\frac{0.619}{0.619+Kn}(1+c_K),
\end{equation}
where $\mu^{at}=1.8 \times 10^{-5}$ Pa s is the viscosity of air at room temperature, $Kn=\bar l / a$ is the Knudsen number, $a$ is the sphere radius, $\bar l = P \bar l_0 / P^{at}$ is the mean free path in air at pressure $P$ while $\bar l_0=66.35$nm is the mean free path in air at atmospheric pressure $P^{at}$ and $c_K=0.31Kn/(0.785 + 1.152Kn+Kn^2)$. In the low pressure limit the mean free path and, therefore, the Knudsen number tend towards infinity, so that $c_K$ tends to zero leaving,
\begin{equation}
\mu \simeq \frac{0.619 \mu^{at}a}{\bar l_0}\frac{P}{P^{at}}=3.56 \times 10^{-7}P
\end{equation}
for a sphere of radius $a=2.2 \mu$m.\\ 
\begin{figure}[H]
\begin{tabular}{cc}
\resizebox{8cm}{!}{\includegraphics{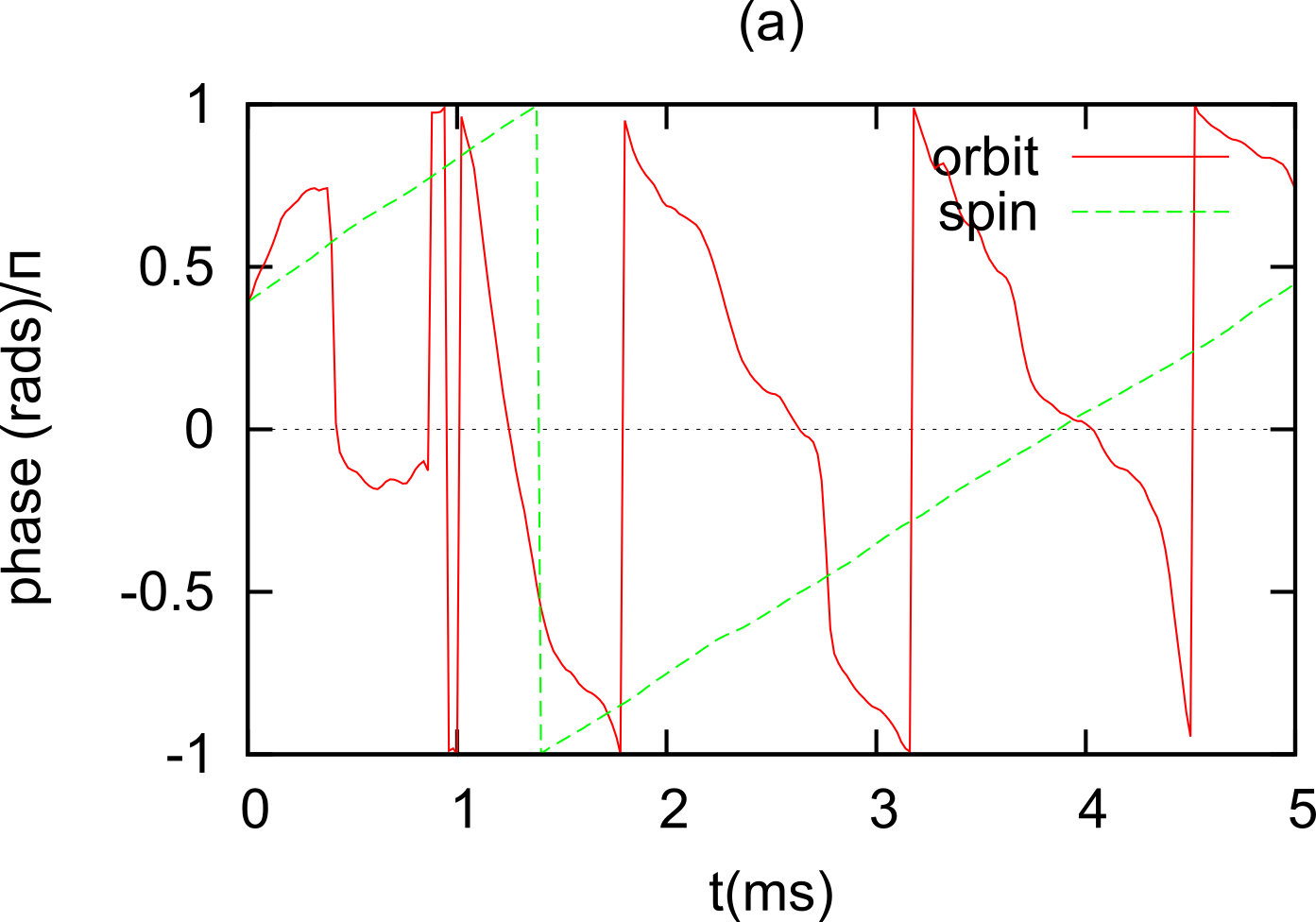}}
\resizebox{8cm}{!}{\includegraphics{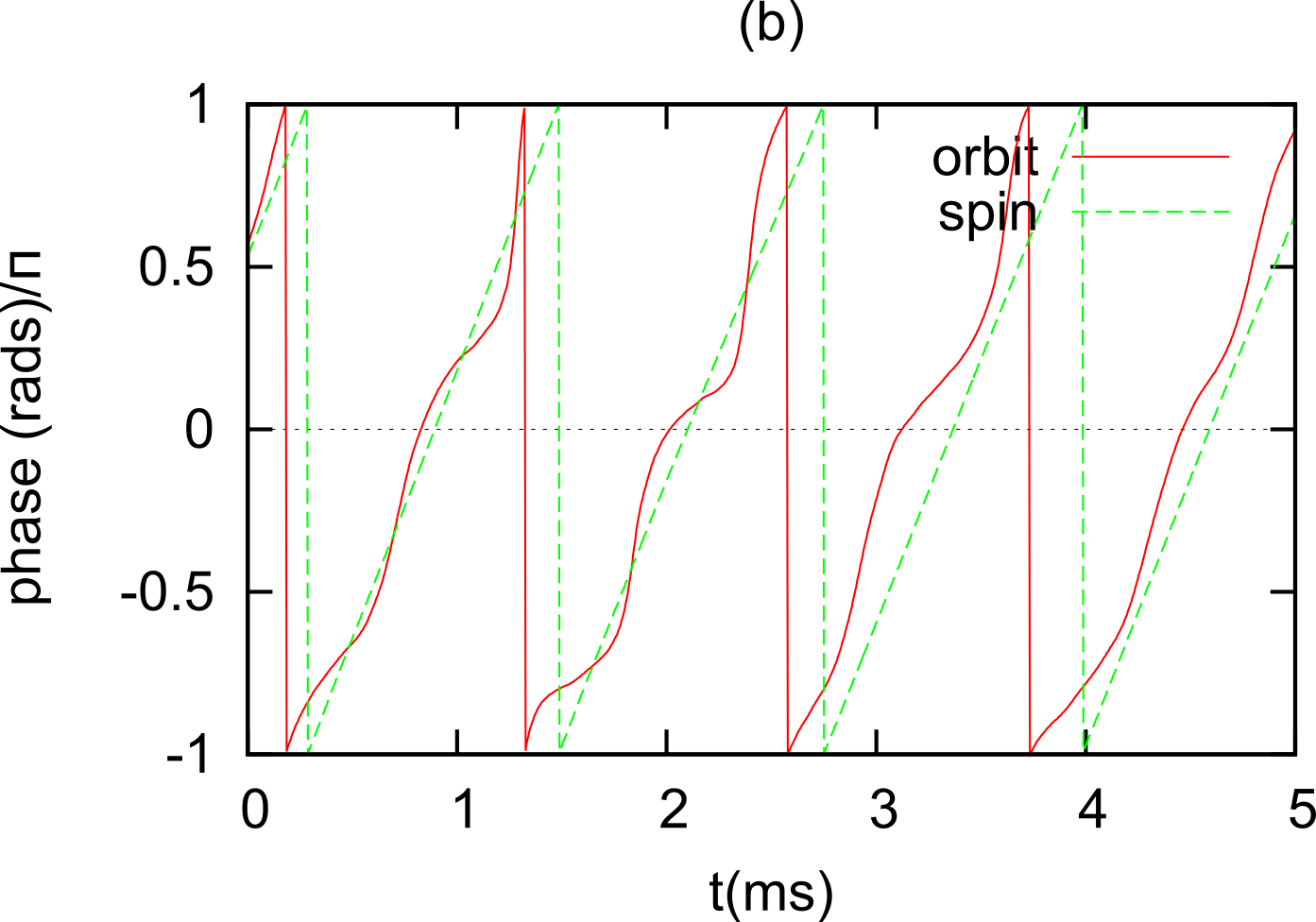}}\\
\resizebox{8cm}{!}{\includegraphics{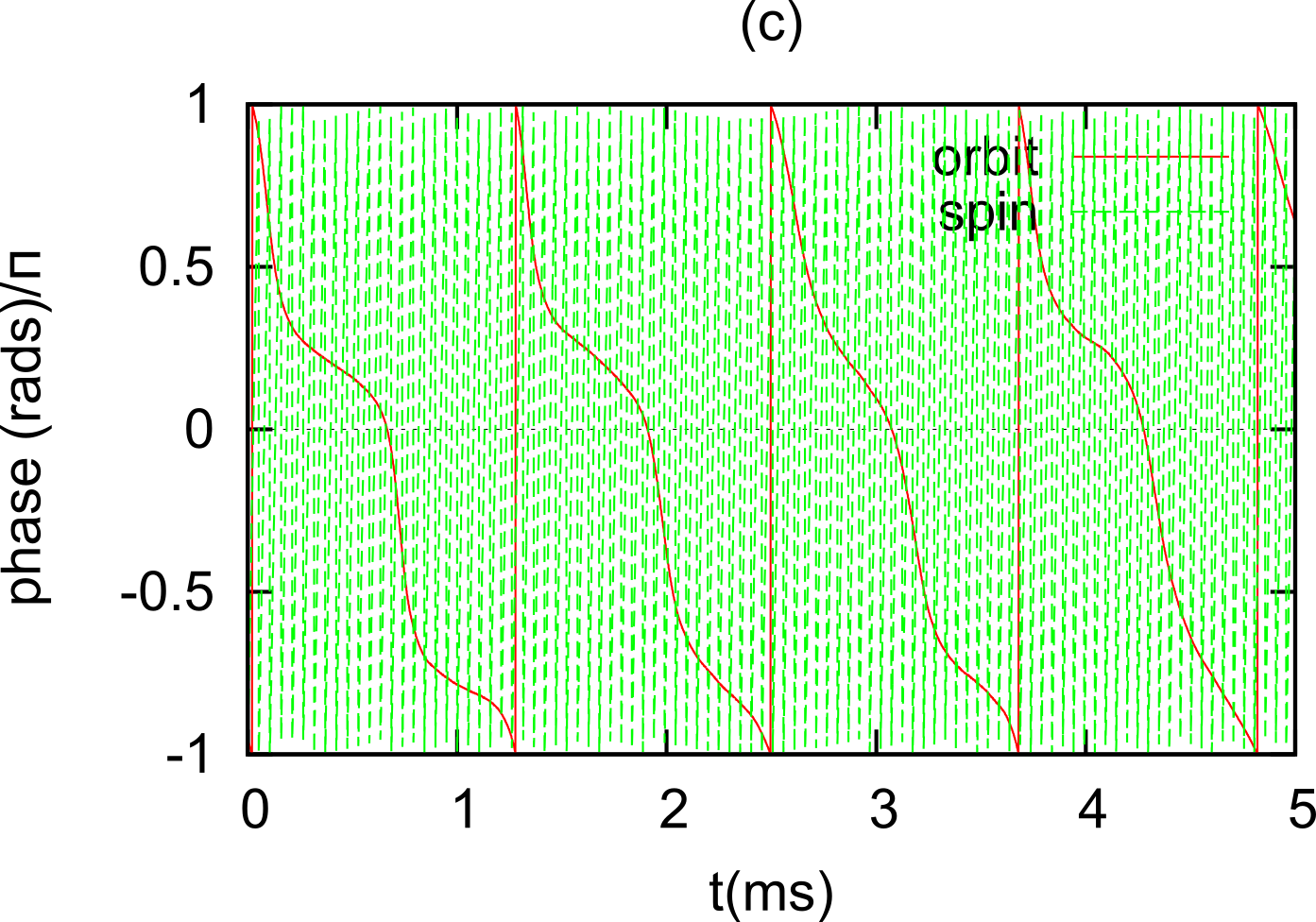}}
\resizebox{8cm}{!}{\includegraphics{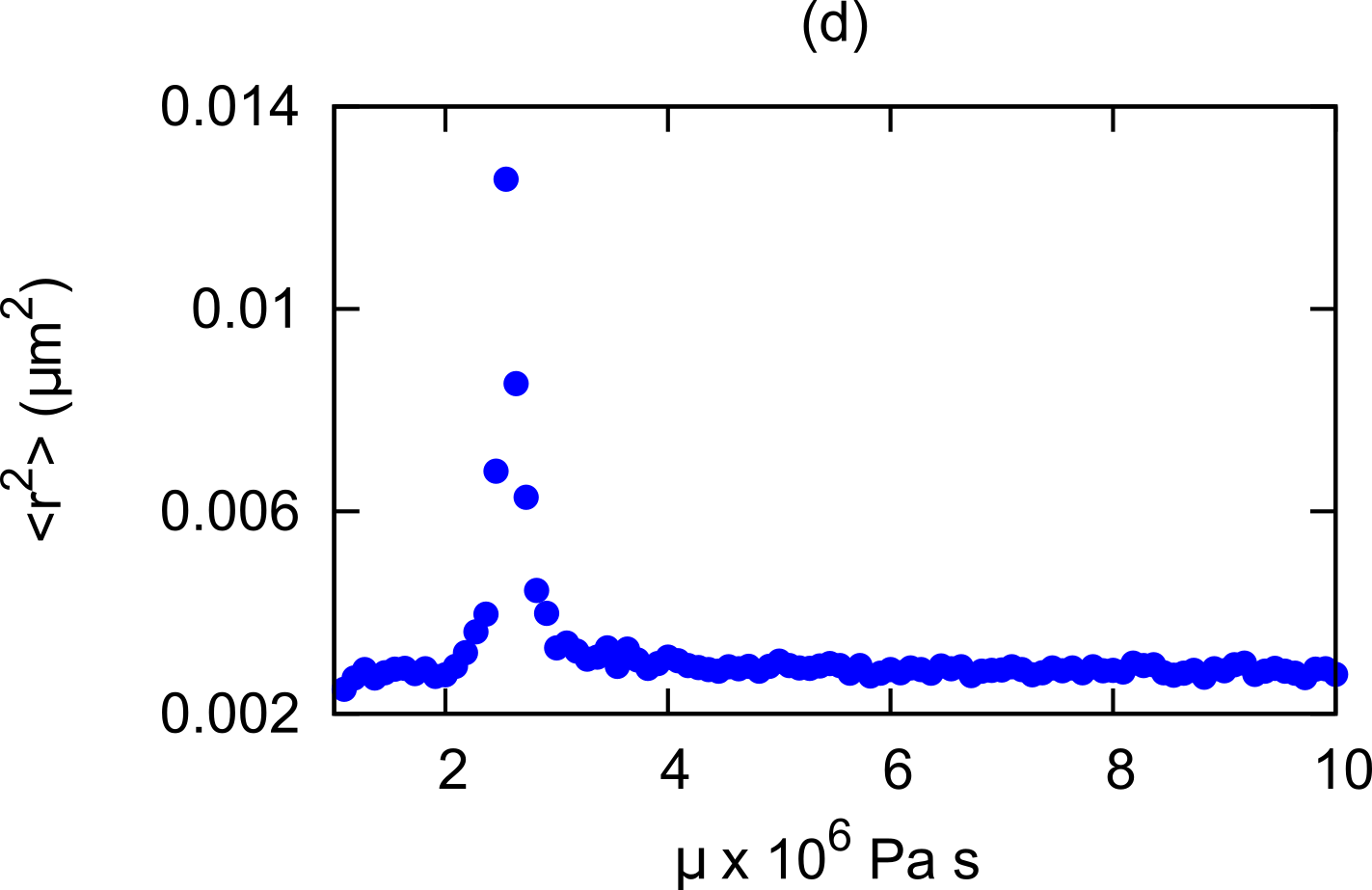}}
\end{tabular}
\caption{\label{fig:phases} Comparison of the orbital phase, $\arctan(y/x)$ with the spin phase, $\phi$ for varying viscosity, (a) $\mu=10^{-7}$Pa s - spin rotation rate much higher than trap frequency, (b) $\mu=2.54 \times 10^{-6}$Pa s - spin rotation rate similar to trap frequency and (c) $\mu=10^{-5}$ Pa s - spin rotation rate slower than trap frequency. (d) $\langle r^2 \rangle$ as a function of viscosity. The sharp peak occurs when the spin frequency matches the trap frequency.}
\end{figure}
\noindent We use the integration scheme described in \cite{sun2008langevin}, which integrates Eq. (\ref{eq:flang}) for general rigid bodies. Optical calculations are performed with T-matrix theory and the forces are calculated through integrals of the optical momentum flux, given by the Maxwell stress tensor, through a closed surface surrounding the particle. The particle itself is modelled as a homogeneous sphere with uniform, birefringent refractive indices \cite{Simpson2009Opt}. It is rotationally symmetric, with its symmetry axis parallel to a vector, $\uvec$. Although this ignores the internal structure of real vaterite particles, it is, nonetheless, a birefringent particle with the same overall symmetry and should be expected to behave in a similar way to a real particle. Numerical simulations are performed with a sphere of nominal parameters corresponding to experimental conditions, i.e. the radius is $a=2.2\mu$m, and the ordinary and extraordinary refractive indices are those of bulk vaterite, i.e. ($n_e=1.65$, $n_o=1.55$) and the density is 2650 kgm$^{-3}$. The modelled beam is a circularly polarized Gaussian beam, rendered by the Richards and Wolf formulation, with the same parameters as those used in the experiment. An optical power of $5mW$ is used for all simulations. As the simulation runs, a Brownian trail is collected.\\
The system has three important time scales. Circular polarization causes the particle to spin, the spin rotation rate reaches a steady state when the optical torque balances the rotational drag i.e. $\tau_z=\xi_r \Omega_s$, with rotational time period $T_s=2\pi\xi_r/\tau_z$. The relaxation time for the velocity of the particle is $T_v=m/xi_t$ and the time period for translational oscillations in the trap is $T_t=2\pi\sqrt{m/k}$, for trap stiffness $k$. Decreasing the pressure, decreases the viscosity and therefore the rotational and translational drag coefficients, $\xi_t$ and $\xi_r$. The time period for spin rotations decreases, $T_s$, while the time period for oscillations in the trap, $T_t$, remains constant. As discussed in the main text, translational oscillations in $x$ and $y$ directions tend to be $\pi/2$ radians out of phase, resulting in stochastic or, for lower pressures, deterministic orbital rotations. In Fig. (\ref{fig:phases}) we plot the orbital phase ($\phi_o=\arctan(y,x)$, the polar angle of the centre of mass) and the phase angle of the spin rotation (given by the angle that the symmetry axis, $\uvec$, makes with the $x$ axis)
 against time for simulations performed at a sequence of decreasing pressures. At higher pressure (panel (a)) the orbital motion is highly stochastic and irregular and, in general, slightly faster than the spin rotation, i.e. $T_s<T_t$. At a lower pressure, the spin and orbital rotation rates are similar, $T_s\sim T_t$ and the two quantities show a loose phase locking, panel (b). In this regime, the angle between $\uvec$ and the vector connecting the beam axis with the centre of mass remains approximately constant as the particle executes its spin and orbit rotations. When this relative orientation maximizes the azimuthal force, a weak instability appears, panel (d). Finally, for lower pressure, the spin rate vastly exceeds the orbital rate, $T_s \gg T_t$, panel (c). \\
 \begin{figure}[H]
\begin{tabular}{cc}
\resizebox{8cm}{!}{\includegraphics{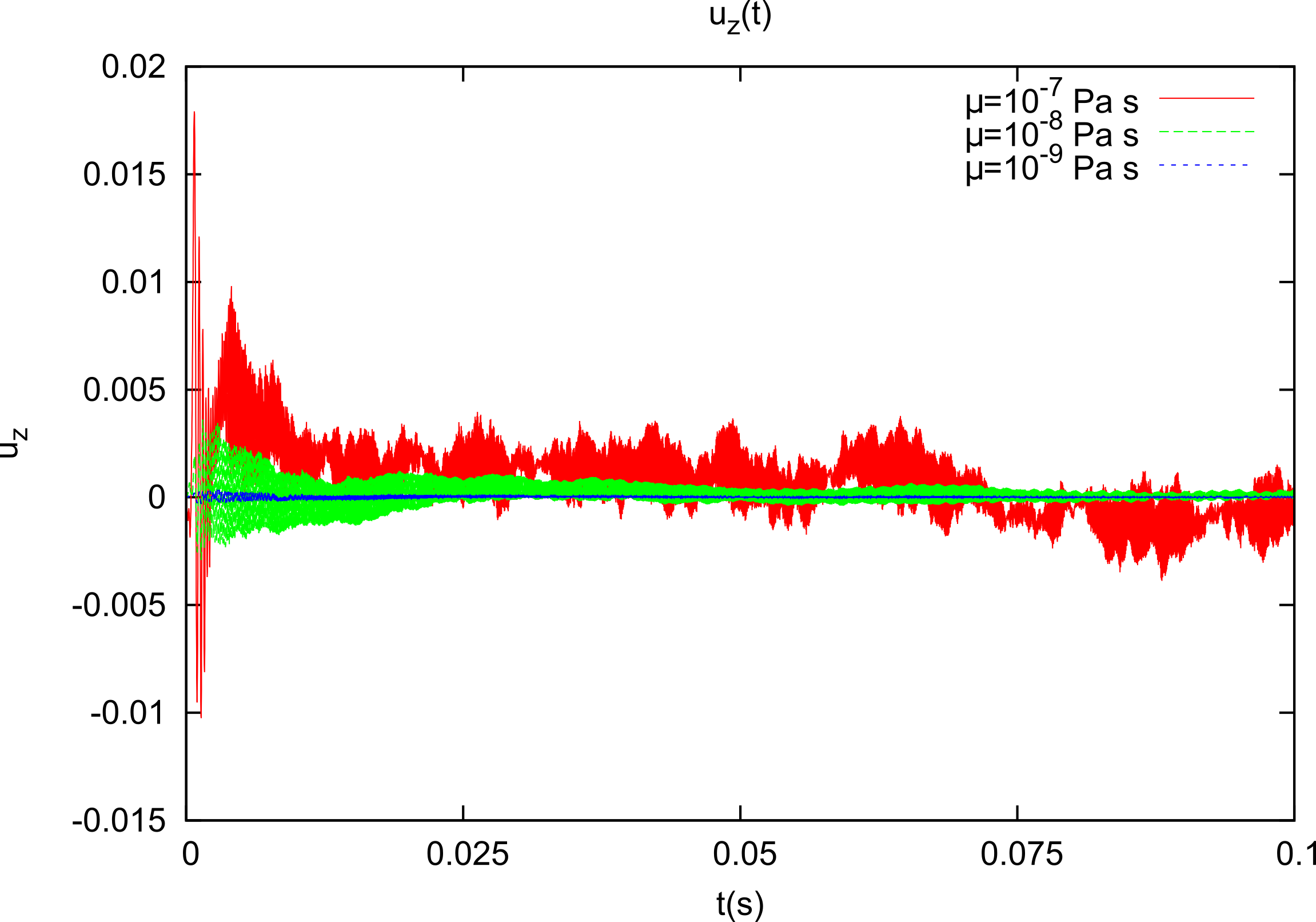}}  
\resizebox{5.5cm}{!}{\includegraphics{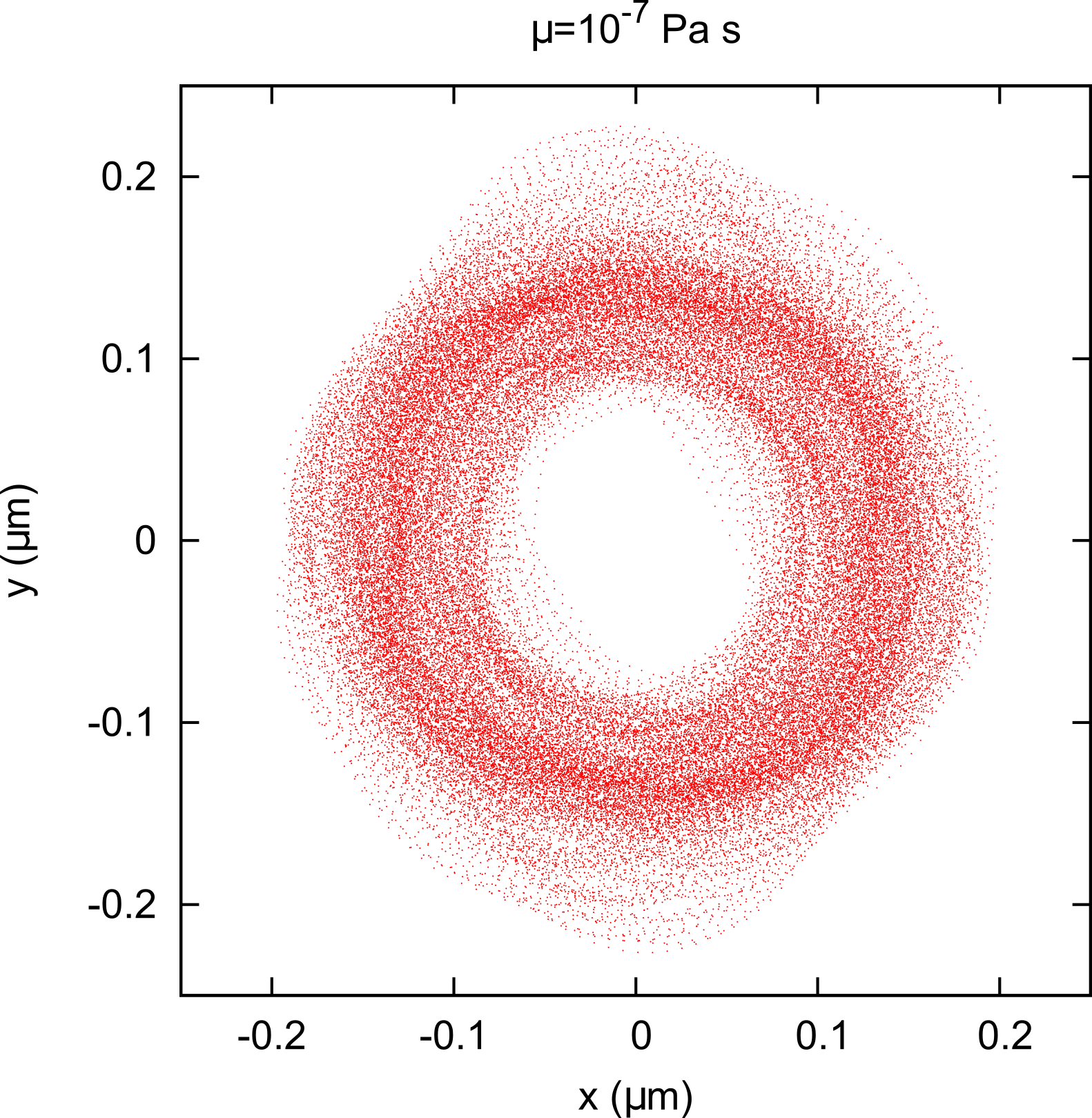}} 
\end{tabular}
\caption{\label{fig:sims_detailed} LHS: vertical component of the symmetry axis, $u_z$ for spinning particles at three different viscosities. RHS: Scatter plot of the centre of mass at $\mu=10^{-7}$ Pa s.}
\end{figure}
 As the pressure is further reduced, the torque required to rotate the symmetry axis, $\uvec$, out of the transverse $xy$ plane increases and the rotation is stabilized, confining $\uvec$ to the transverse plane, Fig. (\ref{fig:sims_detailed}), left hand panel. Noisy limit cycles are formed (right hand panel), as discussed in the main text.\\
 In this low pressure regime, then, the orientation of the particle is constrained, so that its symmetry axis, $\uvec$, is confined to the transverse plane as it rotates about an axis parallel to the beam axis, and noisy limit cycles form. As the particle spins, the transverse forces oscillate with the changing orientation of the particle. This is shown in Fig. (\ref{fig:sims_det2}) for the radial and azimuthal components of the force, for short time intervals, at two different viscosities. Also shown are values for the radial coordinate. As discussed in the main text, the azimuthal force changes sign as it oscillates, leaving a small time average. The radial (or gradient) force also oscillates, but its mean value is approximately proportional to the radial coordinate (and in the opposite direction). Importantly, the centre of mass of the particle does not respond to the rapid force oscillations. Physically, this is due to the ratio of the velocity relaxation time, $T_v$, to the time period of the spin rotation, $T_s$ i.e. $T_s/T_v \propto \mu^2$. This can be seen more directly by considering the influence of a fluctuating force, $f_0+f_1e^{i\Omega_st}$ on the motion of a free spherical particle, subjected to a viscous drag i.e.
 \begin{equation}
 f_0+f_1e^{i\Omega_st}-\xi_t \dot x = m \ddot x
 \end{equation}
 where $\Omega_s$ is the spin frequency, which causes the oscillation in the force. The motion consists of a continuous acceleration in the constant force, $f_0$, combined with an oscillation at frequency, $\Omega_s$. Ignoring transients, the amplitude of the oscillatory motion, $A$, is,
\begin{equation}
A=\frac{f_1}{m}\frac{1}{\sqrt{\Big(\Omega_s^4 + \Omega_s^2 \xi_t^2.
\Big)}}
\end{equation}
Since $\Omega_s=\tau_z/\xi_r\propto 1/\mu$ and the translational drag, $\xi_t \propto \mu$, we have $A\propto \mu^2$ in the low viscosity limit. Inertia strongly suppresses the amplitude response of the particle to increasingly rapid oscillations, whilst the continuous response to the constant term, $f_0$, grows due to decreased drag.

\section{Dynamics with orientationally averaged forces}\label{sec:si_effsims}
\begin{figure}[H]
\begin{tabular}{ccc}
\resizebox{6cm}{!}{\includegraphics{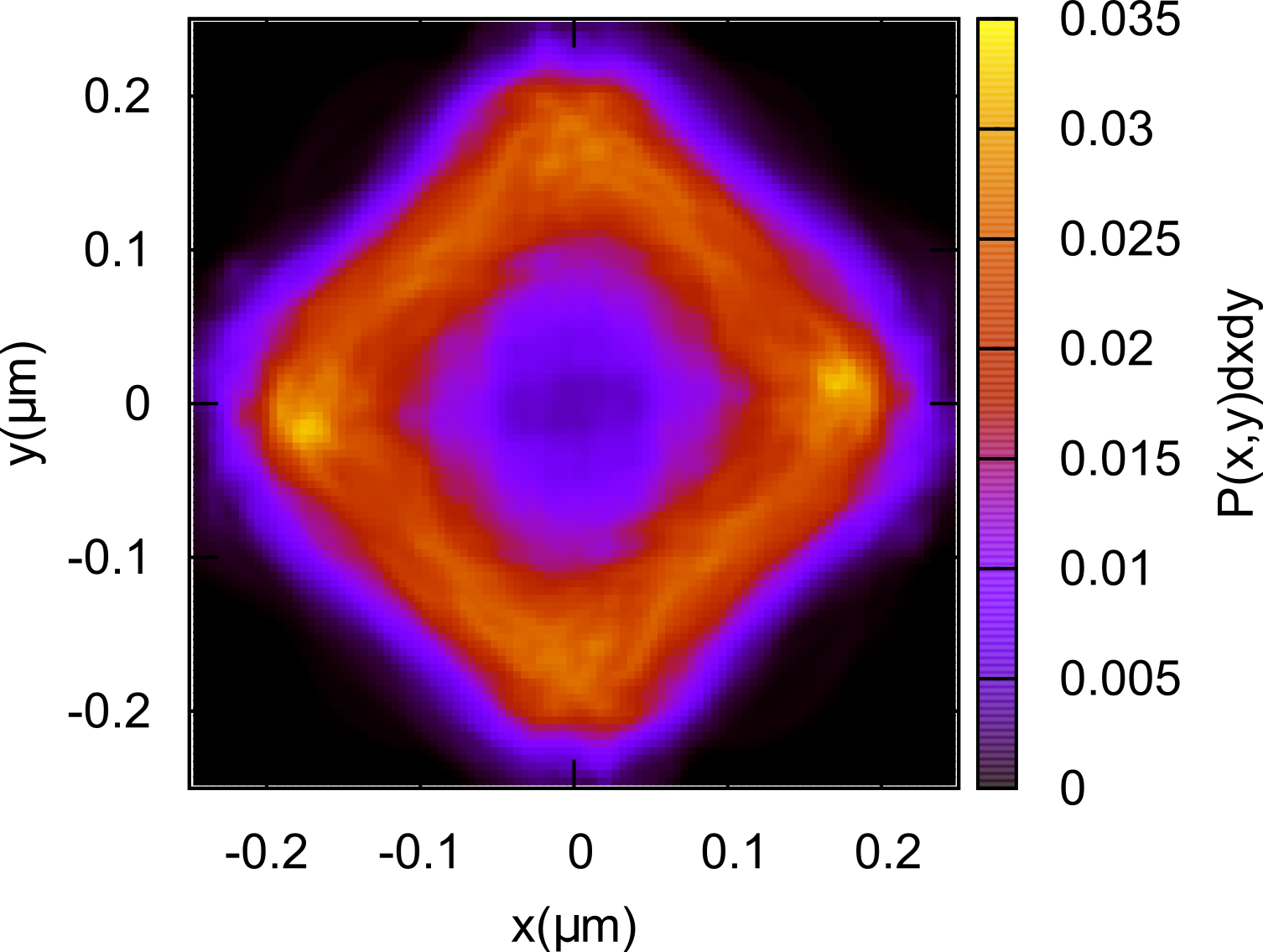}}
\resizebox{6cm}{!}{\includegraphics{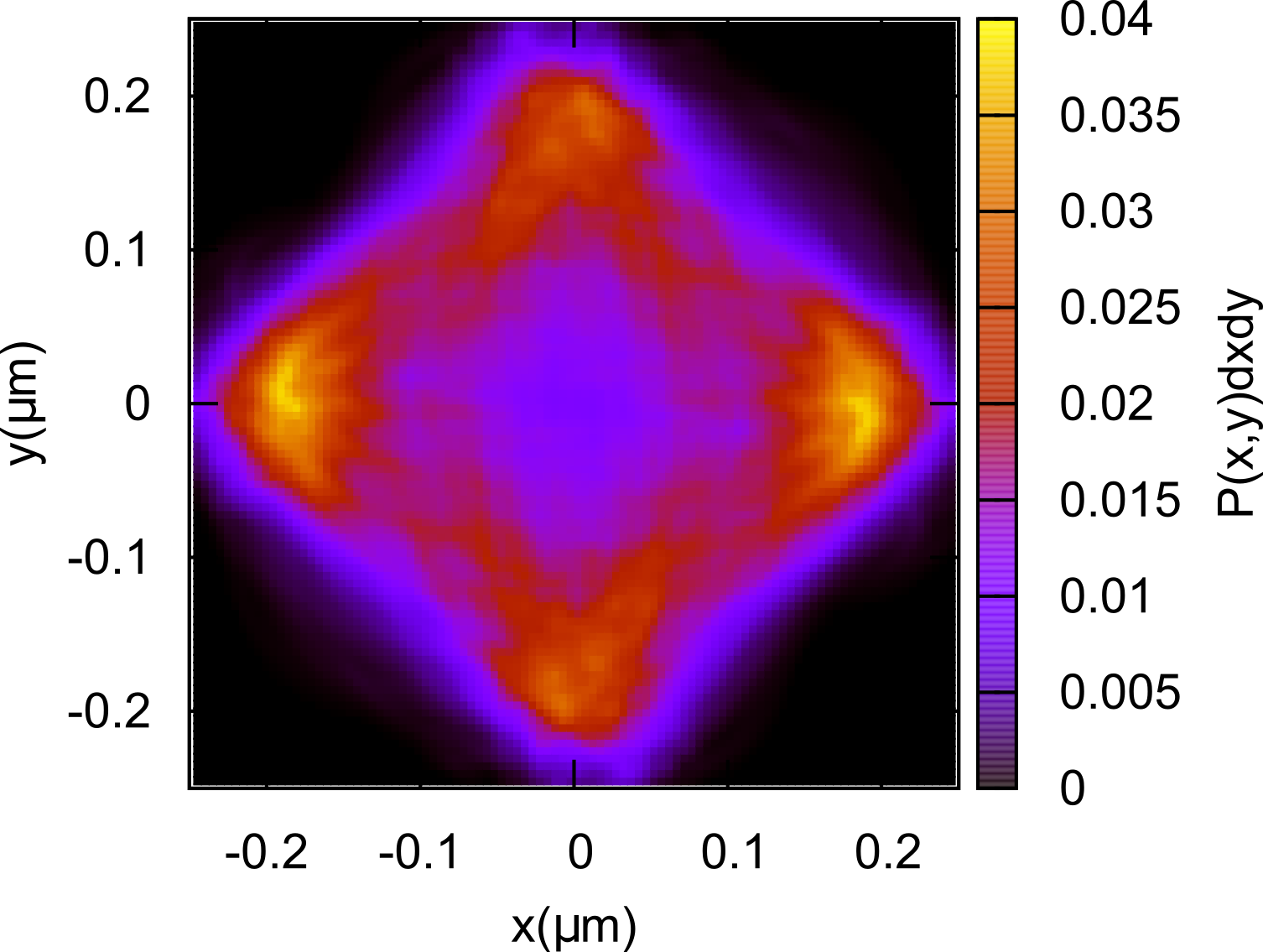}}
\resizebox{6cm}{!}{\includegraphics{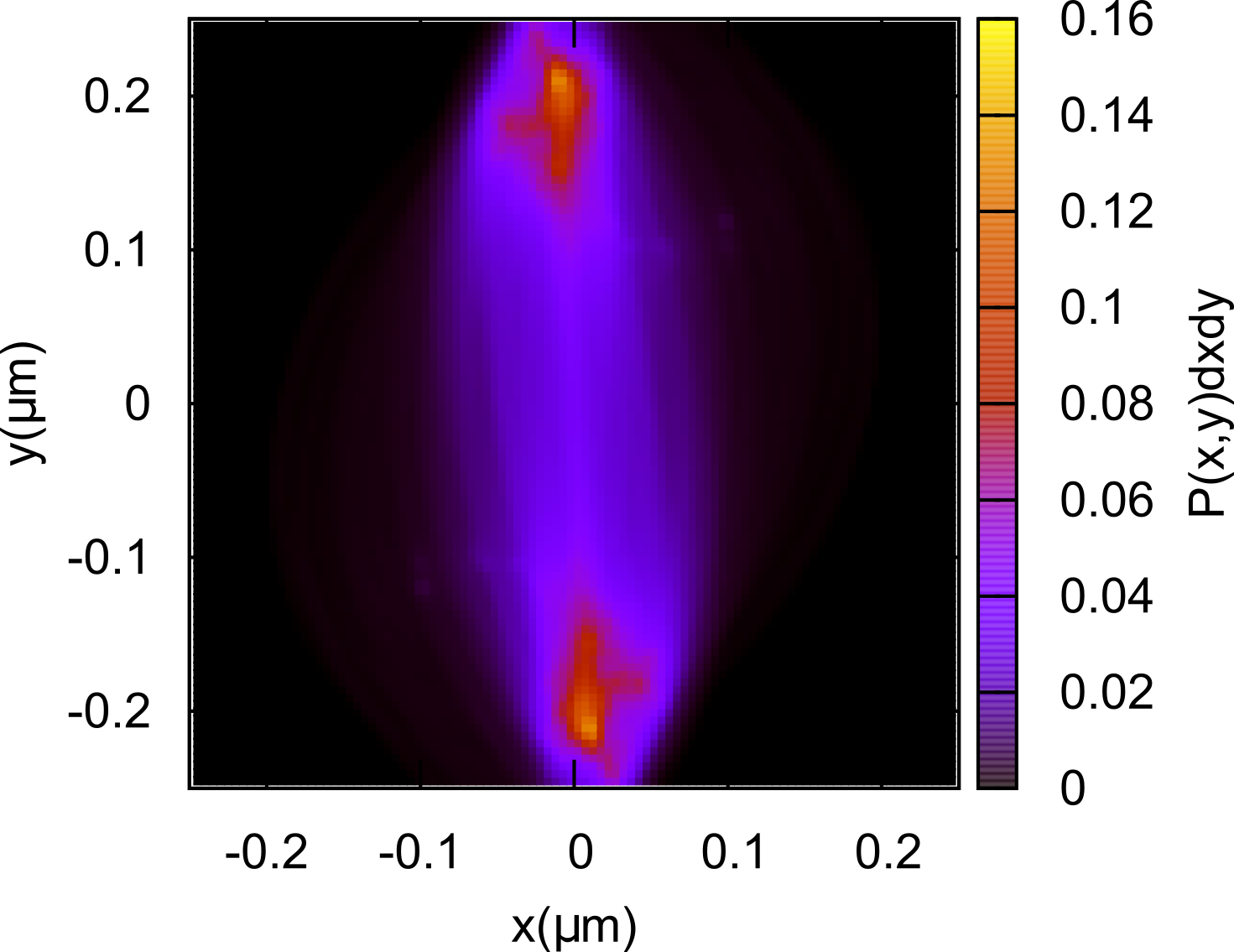}}
\end{tabular}
\caption{\label{fig:densxy} Transverse probability densities for decreasing pressures, $0.3$, $0.03$ and $3 \times 10^{-4}$mbar, from left to right.}
\end{figure}

As described above, the forces acting on a birefringent microsphere depend on its orientation. When it spins very rapidly, these forces oscillate so rapidly that the finite inertia of the particle prevents it from responding significantly. In addition, conservation of angular momentum locks the rotational axis in place, confining the rotating symmetry axis of the particle, $\uvec$, to the transverse plane. Under these conditions, the centre of mass of the particle moves under the influence of orientationally averaged forces, Fig. (\ref{fig:forces})b. This allows us to reduce simulation times by using greater time steps, since the high frequency spinning motion need not be resolved. As shown in Fig. (\ref{fig:forces})b, the direction of the effective azimuthal force reverses when the distance between the beam axis and the centre of mass exceeds $\approx 0.2\mu$m. When centripetal forces propel the particle beyond this radius, the reversal in the azimuthal force acts as a break, reducing the centripetal force and returning the particle to the trap. This process greatly increases the overall stability of the trap, see Fig. (\ref{fig:densxy}). At very low pressures this process results in radial oscillations that are almost confined to a plane (right hand panel, Fig. (\ref{fig:densxy}).

\section{Equilibrium and stability conditions for discrete orbits}\label{sec:si_eqmorb}
\noindent Following \cite{Svak2018Trans}, we give conditions for the stability of circular orbits in a circularly symmetric force field where $f_g(r)$ is the force component in the radial direction and $f_a(r)$ is the azimuthal force.
Neglecting fluctuations, Newton's equations of motion are,
\begin{subequations}
\begin{align}\label{eq:newtoncp}
\dot r &= v \\
m(\dot v - r \Omega^2) &=f_r(r)-\xi_t v \\
m(2 v \Omega + r \dot \Omega) &= f_\phi(r) - \xi_t r \Omega.
\end{align}
\end{subequations}
Here, $r$ is the radial coordinate of the centre of mass, $v$ is the velocity in the radial direction and $\Omega = \dot \phi$ is the angular velocity. Equilibrium conditions correspond to $v=\dot v = \dot \Omega=0$, leaving $-mr\Omega^2=f_r(r)$ (centripetal force balanced by radial force) and $f_\phi(r)=\xi_t r \Omega$ (azimuthal force balanced by drag). Combining these expressions gives $\xi_t^2 = mf^2_\phi(r_o)/rf_r(r_o)$, which determines the viscosity, $\mu$, required to satisfy equilibrium conditions, through $\xi_t=6\pi\mu a$. Stability requires that small perturbations in $r$, $v$ and $\Omega$ result in oscillations that decay with time, returning the particle to the orbit trajectory. Including perturbations, $r=r_o+r_1(t)$, $v=v_1(t)$ and $\Omega=\Omega_o+\Omega_1(t)$  in Eq. (\ref{eq:newtoncp}) and assuming each perturbation has time dependence $\propto e^{\lambda t}$ results in the following secular equation,  

\begin{equation}\label{eq:char}
P(\lambda)=\lambda^3+2\frac{\xi_t}{m}\lambda^2+\frac{\xi_t^2}{m^2}X\lambda+\frac{\xi_t^3}{m^3}Y,
\end{equation}

With:

\begin{subequations}\label{eq:coeff}
\begin{align}
X(r_o) &=1+3\frac{f_r^2(r_o)}{f_\phi^2(r_o)} + r_o\frac{f_r(r_o)f_r'(r_o)}{f_\phi^2(r_o)}, \\
Y(r_o) &= \frac{f_r(r_o)}{f_\phi^2(r_o)}\left[ f_r(r_o)+r_o\left( f_r'(r_o)-2\frac{f_r(r_o)}{f_\phi(r_o)}f_\phi'(r_o) \right) \right].
\end{align}
\end{subequations}
Orbital stability requires that the real parts of each of the roots, $\lambda$, or Eq. (\ref{eq:char}) are negative.\\
Although equilibrium conditions are easily satisfied in a purely linear force field (e.g. with $f_g(r)\propto r$ and $f_a(r)\propto r$), orbital stability requires appropriate curvature. \\
Applying the above criteria to the effective azimuthal force applied to the vaterite microparticle, Fig. (\ref{fig:forces}), the equilibrium and stability conditions can be graphically represented as follows,
\begin{figure}[H]
\begin{tabular}{cc}
\resizebox{9cm}{!}{\includegraphics{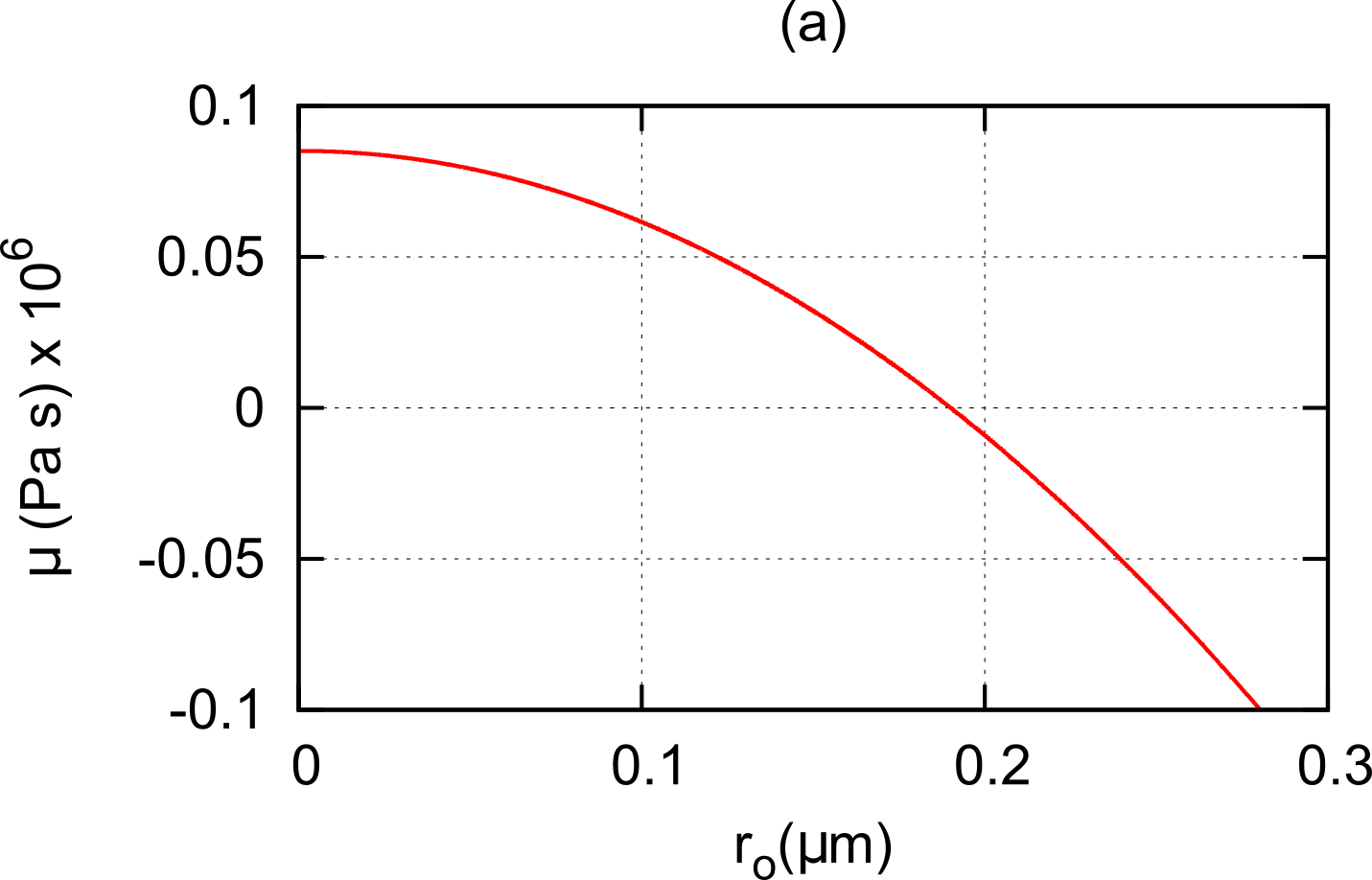}}
\resizebox{9cm}{!}{\includegraphics{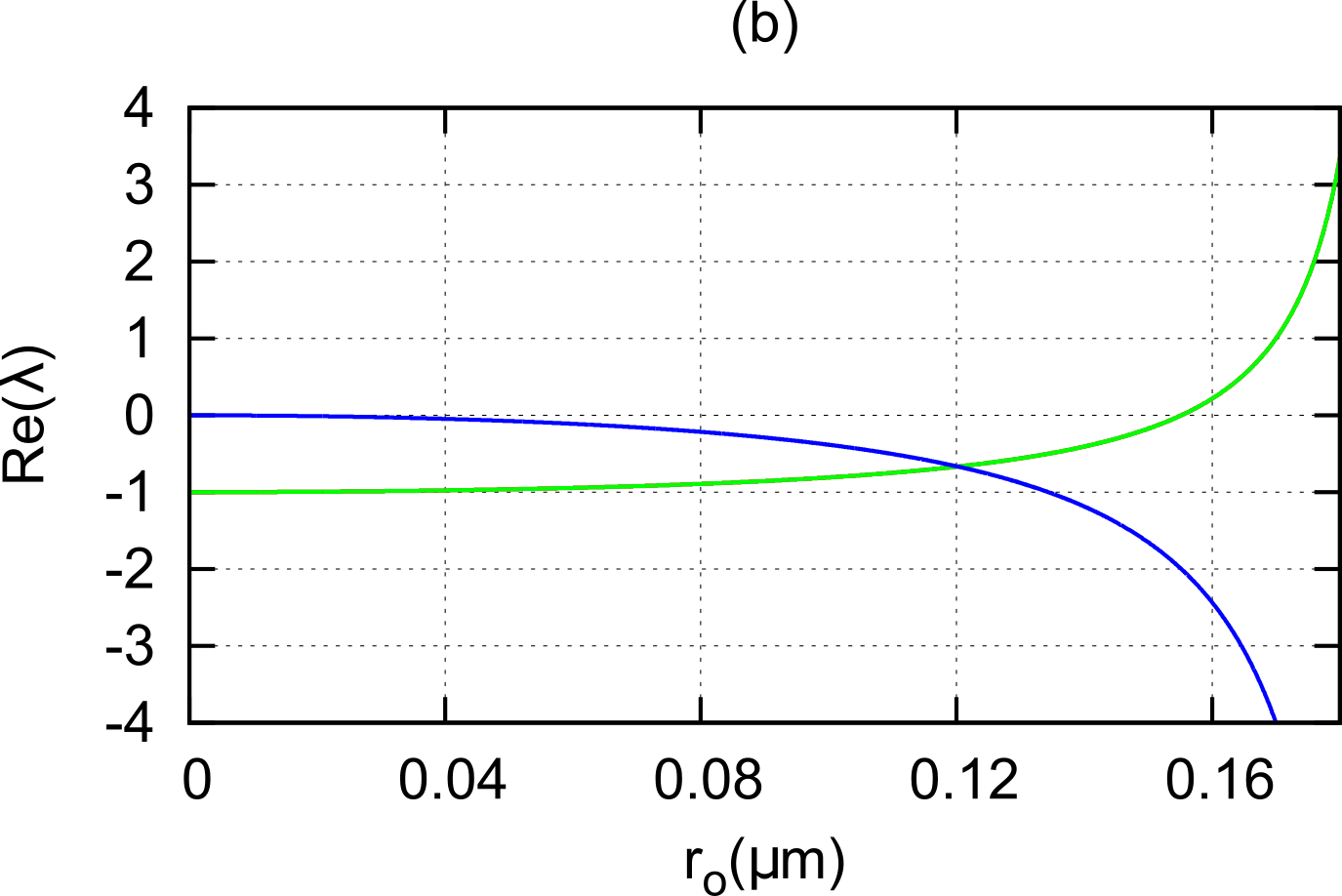}}
\end{tabular}
\caption{\label{fig:stab} (a) The viscosity, $\mu(r_o)$ required to satisfy equilibrium conditions for an orbit of radius $r_o$. (b) Eigenvalues of the secular equation, Eq. (\ref{eq:char}) as a function of orbit radius, $r_o$.}
\end{figure}
\noindent In combination, the graphs in Fig. (\ref{fig:stab}) show that stable orbits can be formed for $r_o \leq \approx 0.15 \mu m$ and $2 \times 10^{-8} \leq \mu \leq 8 \times 10^{-8}$ Pa s. As described in the main text, when used in stochastic simulations these parameters are confirmed to generate fluctuating orbits. Finally, we note that \cite{Svak2018Trans} describes methods to estimate the fluctuations in these stable orbits, the results of which are consistent with the simulations presented here.

\section{Fitting procedure for limit cycles}\label{sec:si_lcfit}
In order to evaluate the limit cycle dimensions, and quantify the fluctuations away from them, we first need to estimate the underlying deterministic path. To do this, we fit a low order Fourier series to the noisy data. The procedure is at follows.\\
\begin{enumerate}
\item Take experimental measured coordinates of the centre of mass, $(X_n,Y_n)$, $n=1,..,N_t$ for times $t_n=n\Delta t$.
\item Find the centre of the distribution, $x_0=\sum_{n=1,..,N} X_n / N_t$, and $y_0=\sum_{n=1,..,N} Y_n / N_t$.
\item Subtract centre from coordinates to give $(x_n,y_n)=((X_n-x_0), (Y_n-y_0))$
\item Translate to circular polar coordinates, $(r_n,\phi_n)$ with $r_n=\sqrt{x^2_n+y^2_n}$ and $\phi_n = \arctan(y_n,x_n)$.
\item Consider the points $(r_n,\phi_n)$ as a distribution in a 2d space, $r,\phi$. Fit a low order Fourier series, 
\begin{equation}
r_o(\phi) = \bar r_o + \sum_{n=1,5} a_n\cos(n\phi)+b_n\sin(n\phi),
\end{equation}
to the noisy data by minimizing the error in the least squares sense.
\end{enumerate}
To quantify fluctuations transverse to the limit cycle we consider the variance in the data after subtracting the fitted limit cycle i.e. the variance in $(r'_n,\phi_n)=((r_n-r(\phi_n)),\phi_n)$.
\section{Phase diffusion for limit cycles, with and without feedback}\label{sec:si_phasediff}
\noindent Below we consider the stochastic motion of a particle moving on a circular path in the underdamped regime. The required Langevin equation in circular polar coordinates is,
\begin{subequations}\label{eq:Lang_cp}
\begin{align}
m(\ddot r - r \dot \phi^2)=f_r(r)-\xi'_t \dot r + f^L_r(t), \label{eq:Lang_cpa}\\
m(2\dot r \dot \phi + r \ddot \phi) = f_\phi(r) - \xi'_t r \dot \phi + f^L_\phi(t).\label{eq:Lang_cpb}
\end{align}
\end{subequations}
Where $\xi'_t$ is a damping coefficient for translational motion, in this case including a feedback induced term i.e. $\xi'_t = \xi^{fb}_t + \xi_t$.  $m$ is the mass, $r$ and $\phi$ are the particle coordinates. The Langevin forces, $f^L_{r,\phi}$ are normalized according to the viscous drag (the drag without the feedback induced contribution) i.e.
\begin{equation}
\langle f^L_{r,\phi}(t) f^L_{r,\phi}(t') \rangle = 2k_BT\xi_t\delta(t-t'),
\end{equation}
where $\xi_t=6\pi\mu a$ is the viscous Stokes drag coefficient.\\
We next assume that the radial coordinate is approximately constant, $r=r_o$, and that radial fluctuations are negligible so that they do not influence the evolution of the azimuthal coordinate, $\phi$, too much. With these assumptions, $\phi$ becomes a reasonable measure of the phase of the oscillator. This allows us to neglect the radial motion described by Eq. (\ref{eq:Lang_cpa}). The equation of motion for $\phi$ is now,
\begin{equation}
\ddot \Omega + \frac{\xi'_t}{m}\Omega = \frac{f_\phi(r) + f^L_\phi(t)}{mr}, \\
\end{equation}
where $\Omega = \dot \phi$. To simplify the algebra we transform to a frame that rotates with the particle i.e. $\Omega \rightarrow \Omega + \Omega_o$ where $\Omega_o=f_\phi / mr_o$, 
\begin{subequations}\label{eq:Lang_phase}
\begin{align}
\dot \Omega + \gamma\Omega &= \Gamma(t), \\
\langle \Gamma(t) \Gamma(t') \rangle &= \frac{2k_BT\xi_t}{m^2r_o^2}\delta(t-t') \equiv A\delta(t-t')
\end{align}
\end{subequations}
with $\gamma=\xi'_t/m$ and $\Gamma(t)=f^L_\phi / mr_o$ and $A$ defined in the final term on the right. Eq. (\ref{eq:Lang_phase}) is identical to the underdamped Langevin equation for a particle in free space with modified drag and normalization of the thermal fluctuations. \\
Following the procedure of formal integration \cite{coffey2012langevin} gives the phase diffusion, 
\begin{equation}
\Big \langle \big(\phi(\Delta t) - \bar \phi(\Delta t) \big)^2 \Big \rangle = \Big(\Omega^2(0)- \frac{A}{2\gamma} \Big) \frac{(1-e^{-\gamma \Delta t})^2}{\gamma^2}+\frac{A}{\gamma^2}\Delta t - \frac{A}{\gamma^3}(1-e^{-\gamma \Delta t}).
\end{equation}
The limit for large $t$ is, 
\begin{equation}
\Big \langle \big(\phi(\Delta t) - \bar \phi(\Delta t) \big)^2 \Big \rangle = \frac{A}{\gamma^2}\Delta t \equiv \frac{2k_BT}{r_o^2}\frac{\xi_t}{(\xi^{fb}_t+\xi_t)^2} \Delta t \sim \frac{2k_BT}{r_o^2}\frac{\xi_t}{(\xi^{fb}_t)^2} \Delta t,
\end{equation}
where the final term corresponds to the limiting behaviour when the feedback induced damping is dominant. In this regime, the rate of phase diffusion decreases with decreasing pressure.
Without feedback, this reduces to the familiar form for phase diffusion about a limit cycle, 
\begin{equation}
\Big \langle \big(\phi(\Delta t) - \bar \phi(\Delta t) \big)^2 \Big \rangle = \frac{2k_BT}{r_o^2}\frac{1}{\xi_t} \Delta t.
\end{equation}
Without feedback, decreasing the pressure increases the rate of phase diffusion, the reverse of the behaviour predicted for phase diffusion with feedback.
\section{Scaling for dipolar particles}\label{sec:si_scaling}
\noindent As described in the main text, the dynamical motion of birefringent particles in circular polarized beams depends on the relative sizes of the optical and viscous forces and the relative time scales of the spinning and orbital motion as well as the relaxation times for the position and velocity. For dipolar particles these quantities are as follows.\\
First, the electric polizability tensor of an anisotropic, dipolar particle is \cite{sihvola1994dielectric}:

\begin{equation}\label{eq:pol}
\alpha_i=\frac{\alpha^{CM}_i}{1-i\frac{2}{3}k^3\alpha^{CM}_i} \approx \alpha^{CM}_i + i\frac{2}{3}k^3(\alpha^{CM}_i)^2,
\end{equation}

\noindent where the denominator ensures that the optical theorem is satisfied. $\alpha_i$ is the i'th eiganvalue, $k=2\pi/\lambda$ is the vacuum wave number and $\alpha^{CM}_i$ is the i'th eigenvalue of the Claussius-Mossotti polarizability, 

\begin{equation}\label{eq:pol0}
\alpha^{CM}_i=\frac{3V(\epsilon_i-\epsilon_0)}{(\epsilon_i+2\epsilon_0)}.
\end{equation}
\noindent $\epsilon_i$ is the i'th eigenvalue of the permittivity tensor with $\epsilon_0$ the permittivity of free space and $V=\frac{4}{3}\pi a^3$ the volume of the spherical particle of radius $a$. Importantly, for non-absorbing media, the real part of the polarizability, Eq. (\ref{eq:pol}) is proportional to the volumne, $V$ (i.e. $\propto a^3$) and the imaginary part is proportional to $V^2$ ($\propto a^6$). For a positively birefringent particle, the eigenvector corresponding to the largest eigenvalue of the polarizability tensor will align with the electric polarization and we take, 
$\alpha_{xx}=\alpha_{zz}\equiv\alpha_o$ and $\alpha_{yy}\equiv\alpha_e$. Rotations of the particle about the $z$ axis leave $\alpha_{zz}$ unchanged. In the transverse plane, axial rotations of the 2 by 2 tensor representing the polarizability in the $xy$ plane are given by,  \\
\begin{equation}\label{eq:pol1}
\alpha = \bar{\alpha} \Imat + \Delta \alpha \Rmat_2(\gamma),
\end{equation}
where $\Delta \alpha$ and $\bar \alpha$ are the anisotropy and mean polarizability respectively,
\begin{eqnarray}\label{eq:pol2}
\Delta \alpha = \frac{1}{2}(\alpha_e - \alpha_o),\\
\bar{\alpha} = \frac{1}{2}(\alpha_e + \alpha_o),
\end{eqnarray}
and $\Rmat_2$ is,
\begin{equation}
\Rmat_2(\gamma) =
\begin{bmatrix}
\cos(2\gamma) & \sin(2\gamma) \\ \sin(2\gamma) & -\cos(2\gamma) 
\end{bmatrix}.
\end{equation}
$\gamma$ is the rotation angle about the $z$ axis. 
\subsection{Spinning motion}
\noindent The equilibrium spin rotation rate is given by the ratio of the axial spin torque, $\tau_z$, to the rotational drag $\xi_r=8\pi\mu a^3$, 

\begin{equation}
\Omega_s=\tau_z/\xi_r.
\end{equation}

\noindent The torque acting on the dipole is, 

\begin{equation}
\tauvec = \frac{1}{2}\Re\Big(\Pvec \times \Evec^{\ast} \Big),
\end{equation}
where $\Pvec$ is the polarization, $\Pvec=\alpha \Evec$, and $\Evec$ is the electric field vector. Combining the above equations gives \cite{simpson2016synchronization},
\begin{equation}\label{eq:tauz}
\tau_z=\frac{1}{2}\chi\Re(\Delta \alpha) + \frac{1}{2}\sigma \Im(\bar \alpha).
\end{equation}
$\chi$ and $\sigma$ are the Stokes parameters,
\begin{eqnarray}\label{eq:stokes}
\chi&=&2\Re(E_yE_x^\ast),\\
\sigma&=&2\Im(E_yE_x^\ast).
\end{eqnarray}
$\chi$ describes the degree of oblique linear polarization, relative to the orientation of the particle (aligned with the coordinate axes) and $\sigma$ is th degree of circular polarization. 
The $z$ component of the torque, $\tau_z$ in Eq. (\ref{eq:tauz}), has two contributions. The first is an alignment torque, twisting the particle to align itself with the preferred polarization. It is proportional to the real part of the anisotropy, $\Re(\Delta \alpha)\propto V$. The second is a spin torque, caused by the angular momentum associated with the circular polarization, $\sigma$. It is proportional to the imaginary part of the real polarizability, $\Im(\bar{\alpha}) \propto V^2$. For a perfect circularly polarized beam, $\chi=0$ and $\sigma=|E_x|^2+|E_y|^2=2w^t_e$, where $w^t_e$ is the energy density in the transverse part of the electric field.\\
Finally, Eqns. (\ref{eq:pol}),(\ref{eq:tauz}) and (\ref{eq:pol1}), (\ref{eq:pol2}) give the scaling of the spin rotation rate for small particles, 
\begin{equation}\label{eq:oms}
\Omega_s = \frac{w^t_e\Im(\bar{\alpha})}{8\pi\mu a^3} \propto \frac{w^t_ea^3}{\mu}.
\end{equation}
Thus, the angular velocity of the spin rotation decreases with decreasing particle size, in the dipole regime and increases with the intensity of the light and with decreasing viscosity, $\mu$ (and therefore pressure).\\\\

\subsection{Forces on the centre of mass}
\noindent The i'th component of the force on a dipolar particle is \cite{chaumet2000time},
\begin{equation}
f_i=\frac{1}{2}\Re(P_j \partial_i E_j^\ast)
\end{equation}
\noindent Restricting attention to the $x$ and $y$ components (the $z$ component being independent), and applying Eq. (\ref{eq:pol1}) gives,
\begin{equation}
f_i=\frac{1}{2} \Re\big[\bar{\alpha}E_j\partial_iE^\ast_j \big]+\frac{1}{2}\Re \Big[\Delta \alpha (\Rmat_2(\gamma) \Evec)_j\partial_i E_j^\ast \Big] \equiv f^{(1)}_i + f^{(2)}_i(\gamma).
\end{equation}
\noindent The first term on the right is a rotationally averaged force, the second is dependent on orientation and averages to zero when integrated over $-\pi \leq \gamma \leq \pi$. Considering the first term $f_i^{(1)}$, and separating $\bar{\alpha}$ into its real and imaginary parts gives,
\begin{subequations}\label{eq:dfrc}
\begin{align}
f_i^{(1)}=&\frac{1}{2}\Re(\bar{\alpha})\Re \big[E_j\partial_iE_j^\ast \big] - \frac{1}{2}\Im(\bar{\alpha})
Im\big[E_j\partial_iE_j^\ast \big] \\
=&\frac{1}{2}\Re(\bar{\alpha})\partial_i w_e + \Im(\bar{\alpha}_i)\omega\pvec^o_i \label{eq:dfrc2},
\end{align}
\end{subequations}
\noindent where $\omega$ is the optical frequency, $w_e$ is the electric energy density and $p^o$ is the canonical momentum. Thus, the first term in Eq. (\ref{eq:dfrc2}) is the (orientationally averaged or \textit{effective}) gradient force and the second term is an effective scattering force, or a radiation pressure force, proportional to the canonical momentum, $p^o$ which is independent of spin, $\sigma$. In a circularly polarized beam, the gradient force is directed towards the beam axis as usual and can be recognised as the radial force, $f_r$ in the main text. Components of $p^o$ swirl about the beam axis, due to helical inclination of the wavefronts \cite{bekshaev2011internal}, generating the azimuthal force, $f_\phi$. Thus, in accordance with Eq. (\ref{eq:pol}), the gradient force is $\propto V\sim a^3$ and the azimuthal force on the dipole is $\propto V^2 \sim a^6$. In the small particle limit, the gradient force completely dominates, and the azimuthal force is negligible,
 
\begin{equation}
\frac{f_\phi}{f_r} = \frac{2\Im(\bar{\alpha})}{\Re(\bar{\alpha})\partial_iw_e} \propto a^3
\end{equation}

\noindent The orientation dependent part of the force, $f^{(2)}(\gamma)$ can be expanded as,

\begin{equation}
f^{(2)}_i(\gamma)=\frac{1}{2}\Re\Big[\Delta \alpha\Big(\big(E_x\partial_iE^\ast_x - E_y\partial_iE^\ast_y \big)\cos(2\gamma)+\big(E_y\partial_iE^\ast_x+E_x\partial_iE^\ast_y \big)\sin(2\gamma) \Big) \Big].
\end{equation}

\noindent For a perfect, circularly polarized beam with $E_x=f(\rvec)$ and $E_y=if(\rvec)$ for some function $f(\rvec)$, $f^{(2)}_i(\gamma)$ vanishes identically. When the beam is less than perfect, $f^{(2)}_i$ remains small, and its orientational average is zero.   \\
In summary, the forces and the spin torque on the anisotropic dipole considered above turn out to be independent of anisotropy. This is a consequence of the fact that there are two sources of mechanical force and torque on a dipole. The first is related to energy - the total electromagnetic potential energy of the system can be reduced by the particle moving to regions of high intensity. The second is connected with momentum. For dipoles, only the canonical momentum is relevant \cite{bliokh2014extraordinary}, and this is independent of spin. There is one additional form of torque acting on the dipole, and not considered above. This is the alignment torque that orients the particle with respect to the plane of polarization. This torque can be also be thought of as being connected with energy minimization. For example, alignment torques also operate in electrostatic fields. In the small particle limit we see that azimuthal forces become negligible in comparison to gradient forces, stabilizing the centre of mass motion. However, the spin rate, $\Omega_s$, becomes small. Rapid spinning will always be accompanied, therefore, by azimuthal forces. 

\end{document}